\documentclass[12pt,a4paper,final]{iopart}
\usepackage{iopams}
\expandafter\let\csname equation*\endcsname\relax
\expandafter\let\csname endequation*\endcsname\relax
\usepackage{amsmath,amsthm,amssymb,amsfonts} 
\usepackage{graphicx}
\usepackage[utf8]{inputenc} 
\usepackage[usenames,dvipsnames]{xcolor}
\usepackage{float}
\usepackage{tikz}
\usepackage{bm}
\usepackage[caption=false]{subfig}
\usepackage{indentfirst}
\usepackage{dsfont}
\usepackage[breaklinks=true,colorlinks=true,linkcolor=blue,urlcolor=blue,citecolor=blue]{hyperref}
\usepackage{array}
\usepackage{eso-pic}

\newcolumntype{M}[1]{>{\arraybackslash}m{#1}}

\DeclareSymbolFont{matha}{OML}{txmi}{m}{it}
\DeclareMathSymbol{\vary}{\mathord}{matha}{121}

\newcommand{\YZ}{\color{black}}
\newcommand{\bb}{\color{black}}

\usetikzlibrary{arrows.meta}
\newtheorem*{algorithm}{SBD Algorithm}

\begin{document}

\AddToShipoutPictureBG*{%
  \AtPageUpperLeft{%
    \hspace{\paperwidth}%
    \raisebox{-\baselineskip}{%
      \makebox[0pt][r]{Nonlinearity 31(1):R1-R23 (2018)}
}}}%

\title[]{{\bb Identical synchronization of nonidentical oscillators:} when only birds of different feathers flock together}

\author{Yuanzhao Zhang$^{1}$ and Adilson E. Motter$^{1,2}$}
\address{$^1$Department of Physics and Astronomy, Northwestern University, Evanston, Illinois 60208, USA}
\address{$^2$Northwestern Institute on Complex Systems, Northwestern University, Evanston, Illinois 60208, USA}
\eads{\mailto{motter@northwestern.edu}}

\begin{abstract}
An outstanding problem in the study of networks of heterogeneous dynamical units concerns the development of rigorous methods to  probe the stability of synchronous states when the differences between the units are not small. Here, we address this problem by presenting a 
generalization of the master stability formalism  that can be applied to heterogeneous oscillators with large mismatches. Our approach is based on the simultaneous block diagonalization of the matrix terms in the variational equation, and {\bb it leads to} 
dimension reduction that simplifies the original equation significantly. This new formalism 
{\bb allows}
 the systematic investigation 
{\bb of scenarios} in which the oscillators need to be nonidentical in order 
to reach {\bb an identical state, where all oscillators are completely synchronized.}
In the case of networks of identically coupled oscillators, this
 corresponds to breaking the symmetry of the system  
 {\bb as a means} to preserve the symmetry of the dynamical state---{\bb a recently discovered effect termed}   asymmetry-induced  synchronization ({\it AISync}).
{\bb Our framework}
enables us to 
{\bb identify}
communication delay as a new and potentially common mechanism giving rise to {\it AISync}, which we demonstrate using networks of delay-coupled  Stuart-Landau oscillators. 
{\bb %
The results also have potential 
implications for control, as they reveal oscillator heterogeneity as an 
attribute that may be manipulated to enhance the stability of synchronous states.}
\end{abstract}

%
%
\vspace{2pc}

\section{Introduction}

The study of synchronization phenomena in networks of coupled dynamical systems has traditionally focused on either the partial synchronization of nonidentical oscillators, such as in the Kuramoto model \cite{kuramoto1975self},  or the complete synchronization of identical ones,  as in the Pecora-Carroll model \cite{pecora1990synchronization,pecora1998master}. The first concerns primarily studies in the limit of large population sizes and 
{\bb uses} approaches that stem from statistical physics, while the second emphasizes the study of finite-size systems using dynamical systems methods to characterize the stability of synchronous states \cite{abrams2016introduction}. Until recently, little attention 
{\bb was}
given to the possibility of complete synchronization of nonidentical oscillators. This was the case because, on the one hand, there has been {\bb a} lack of rigorous dynamical systems approaches that can be used to study complete synchronization in networks of nonidentical 
{\bb oscillators;}
on the other hand, it was not appreciated that complete synchronization could occur for nonidentical oscillators, let alone that it would lead to interesting new effects. The latter has changed with the recent discovery of so-called asymmetry-induced synchronization ({\it AISync})
{\bb \cite{PhysRevLett.117.114101,zhang2017asymmetry},} 
where complete synchronization becomes stable in networks of nonidentical oscillators because of (not despite) the differences 
between 
the oscillators.  This was demonstrated for networks of identically coupled oscillators, meaning that the symmetry of the system had to be broken to preserve the symmetry of the stable solution---a property that corresponds to the converse of symmetry breaking (hence of chimera states \cite{panaggio2015chimera}) and after which the effect is named.

Motivated by that discovery, in this article we first  present a rigorous framework to analyze complete synchronization 
{\bb in the most general class of coupled nonidentical oscillators that permits complete synchronization, and}
then apply this formalism to characterize a new mechanism through which {\it AISync} can occur. 
{\bb This class includes networks of nonidentical oscillators with arbitrary differences, 
provided that they admit at least one common orbit when coupled.}
Our framework consists of a 
generalization of the master stability function (MSF) formalism  \cite{pecora1998master},  which can be applied to 
{\bb this} class of nonidentical oscillators and several forms of coupling.
 The new mechanism for {\it AISync} identified here is mediated by delay-coupling and is demonstrated for networks of Stuart-Landau oscillators.

Complete (or identical) synchronization refers to the scenario in which all oscillators converge to the same dynamical state (with respect to all of their variables). In a network of $N$ oscillators, where the $d$-dimensional state of the $i$-th oscillator is denoted $\bm{x}_i$, 
complete synchronization corresponds to 
{\bb \begin{equation}
\bm{x}_1(t) = \bm{x}_2 (t) = \cdots = \bm{x}_N(t) \equiv {\bm s}(t) 
\label{eq:sync}
\end{equation}
for all $t$, where ${\bm s}(t)$ denotes the 
synchronous state \footnote{For notational simplicity, throughout the text (but not in equations) the synchronization orbit of individual oscillators ${\bm s}(t)$ will also be used to denote the synchronization orbit of the full network, as a short for the $N\times d$-dimensional vector $({\bm s}(t), \cdots, {\bm s}(t))$}.}
 This should be contrasted with cases in which a condition of the form~\eqref{eq:sync} is satisfied for only some of the variables or a function of the variables, as in the 
{\bb cases}
of 
{\bb identical-frequency (but not identical-phase)} 
synchronization in power-grid networks \cite{motter2013spontaneous,dorfler2013synchronization,rohden2012self} and 
{\bb output-function synchronization} 
in output consensus dynamics \cite{seyboth2015robust,lunze2012synchronization}.

It is instructive to first recall the previous main mechanisms through which {\it AISync} has been demonstrated:
\begin{itemize}

\item Amplitude-dependent coupling in 
{\bb networks of phase-amplitude oscillators} 
  \cite{PhysRevLett.117.114101}, where suitable heterogeneity in the amplitude term stabilizes the otherwise unstable state of complete synchronization.

\item Subnode coupling in multilayer networks \cite{zhang2017asymmetry}, where heterogeneity
{\bb  is required}
 in the internal couplings between different variables of the oscillators in order to stabilize 
 complete synchronization.

\end{itemize}
In the mechanism considered here, on the other hand, the oscillator heterogeneity is in the angular term of 
{\bb delay-coupled}
Stuart-Landau oscillators. This heterogeneity stabilizes the complete synchronization state that would otherwise become unstable in the presence of coupling delay. While we exemplify our results on a selection of 
{\bb representative}
networks, 
the oscillator model {\bb we consider} 
can be tested for {\it AISync} (using our 
formalism)  in any network of identically coupled nodes, {\bb which includes the rich class of vertex-transitive graphs}.
{\bb We also demonstrate
the analog of {\it AISync} in a broader class of networks by showing that
oscillator heterogeneity can stabilize synchronization 
when the oscillators are not necessarily identically coupled.}
For the oscillators that {\bb we} explicitly consider, 
{\bb the latter includes arbitrary regular graphs.}

The 
generalization of the MSF formalism presented  in this work applies to oscillators that are not necessarily similar to each other. 
This should be contrasted with previous 
generalizations of the MSF formalism to systems with small parameter mismatches in the oscillators \cite{sun2009master,acharyya2012synchronization} and systems with small mismatches in the oscillators and coupling functions \cite{sorrentino2011analysis}, where
the focus is on approximate (rather than complete) synchronization. 
{\bb Other approaches, such as the dichotomy technique used in Ref.~\cite{pereira2014towards}, 
are also designed for
approximate synchronization of nearly identical oscillators.}
{\bb Here, while}
we consider oscillators that can differ by more than a small mismatch, our focus is on the case of complete synchronization. A  notable exception in the existing literature to also have considered complete synchronization in a non-perturbative {\bb parameter} regime comes from the control community \cite{zhao2011synchronization}, where  it has been shown that sufficient conditions for the global stability of a state of complete synchronization among nonidentical oscillators can be given based on a Lyapunov function approach. Those conditions are expressed 
{\bb through equations with}
the  dimension of the individual 
{\bb oscillators,} but their verification requires finding 
{\bb time-varying matrices that satisfy}
matrix inequalities for all $t$; moreover, like other Lyapunov function 
{\bb methods,} such an approach has limitations when applied to multi-stable systems. The approach we present, 
on the other hand, gives verifiable necessary and sufficient conditions for the linear stability of 
{\bb states of} 
complete synchronization 
in 
{\bb networks}
with any number of stable states or attractors (including chaotic ones).


The article is organized as follows. 
In Sec.~\ref{sec:msf}, we first develop our framework for nonidentical oscillators in the context of Laplacian-matrix (diffusive) coupling (Sec.~\ref{sec:ndc}). We then discuss the conditions under which the framework also applies to
{\bb two classes of} non-diffusively coupled systems, namely networks with adjacency-matrix coupling (Sec.~\ref{sec:nam}) and networks {\bb with delay coupling} (Sec.~\ref{sec:nwd}).
{\YZ Table~\ref{tbl:tbl1} summarizes the conditions for nonidentical oscillators to admit complete synchronization for each of the three types of couplings we consider.}
In Sec.~\ref{sec:sbd}, we elaborate on the theoretical background and algorithmic implementation of our approach, 
{\bb which is} based on {\bb the irreducible decomposition of}
{\bb an algebraic structure known as}
matrix $*$-algebra.
In Sec.~\ref{sec:ais}, we present our application of the formalism to {\bb establish} 
{\bb networks of delay-coupled Stuart-Landau oscillators as}
 a new class of systems that exhibit {\it AISync}. To that end, 
{\bb following a brief discussion of the delay-coupled dynamics}
 (Sec.~\ref{sec:ind}), 
 {\bb we show} 
 that oscillator heterogeneity can stabilize 
 {\bb an}
 otherwise unstable state of complete synchronization on  representative networks (Sec.~\ref{sec:nind}).  We also demonstrate the 
 {\bb analogs}
 of {\it AISync} for 
networks in which  {\bb the} oscillators are not identically coupled (Sec.~\ref{sec:nonsym})  {\bb and for networks  with unrestricted oscillator parameters (Sec.~\ref{sec:genr})}.
We show  that delay is a key ingredient leading to this effect in 
{\bb the  class of systems we consider}, which suggests that {\it AISync} 
{\bb may be common}
in physical systems, where delay is often significant.  Concluding remarks are presented in Sec.~\ref{sec:concl}.

\begin{table}[t] 
\caption{\YZ Necessary and sufficient conditions for the existence of a complete synchronization state $\bm{s}(t)$. 
Here, $\bm{F}_i$ is the intrinsic dynamics of the $i$-th oscillator, $\bm{H}$ is the interaction function, $\mu_i$ is the indegree of the $i$-th node (denoted by $\mu$ when equal for all nodes), $\sigma$ is the coupling strength, and $\tau$ is the communication delay.
In each case, the conditions are to be satisfied for all $t$.}
\begin{center}
\begin{tabular}{M{2.4cm} || M{7cm} | M{5.6cm} }
    \firsthline \firsthline
    {\small Coupling type} & {\small Networks with arbitrary indegrees} & {\small Networks with common indegrees}
    \\ \hline \hline
    {\scriptsize Laplacian-matrix \vspace{-2mm} \newline coupling (Sec.~\ref{sec:ndc})} &
    {\scriptsize $\bm{F}_i(\bm{s}(t))$ independent of $i$} \newline 
    {\scriptsize $\dot{\bm{s}}(t)\! =\! \bm{F}_i(\bm{s}(t))$}
    & {\scriptsize $\bm{F}_i(\bm{s}(t))$ independent of $i$ \newline
      $\dot{\bm{s}}(t) \!=\! \bm{F}_i(\bm{s}(t))$}
	\\ \hline
    {\scriptsize Adjacency-matrix \vspace{-2mm} \newline coupling (Sec.~\ref{sec:nam})} &
    {\scriptsize $\bm{F}_i(\bm{s}(t)) + \sigma \mu_i \bm{H}(\bm{s}(t))$ independent of $i$ \newline
    $\dot{\bm{s}}(t)\! =\! \bm{F}_i(\bm{s}(t)) + \sigma \mu_i \bm{H}(\bm{s}(t))$}
    & {\scriptsize $\bm{F}_i(\bm{s}(t))$ independent of $i$ \newline 
      $\dot{\bm{s}}(t)\! =\! \bm{F}_i(\bm{s}(t)) + \sigma \mu \bm{H}(\bm{s}(t))$} 
    \\ \hline
    {\scriptsize Delay coupling \vspace{-2mm}
    \newline (Sec.~\ref{sec:nwd})} &
    {\scriptsize $\bm{F}_i(\bm{s}(t)) + \sigma \mu_i \left[ \bm{H}(\bm{s}(t\!-\! \tau)) \!-\! \bm{H}(\bm{s}(t)) \right]\!$ independent of $i$\newline
    $\dot{\bm{s}}(t) \!=\! \bm{F}_i(\bm{s}(t)) + \sigma \mu_i \left[ \bm{H}(\bm{s}(t\!-\! \tau)) \!-\! \bm{H}(\bm{s}(t)) \right]$}
    & {\scriptsize $\bm{F}_i(\bm{s}(t))$ independent of $i$ \newline 
      $\dot{\bm{s}}(t) \!=\! \bm{F}_i(\bm{s}(t)) + \sigma \mu \left[ \bm{H}(\bm{s}(t\!-\! \tau)) \!-\! \bm{H}(\bm{s}(t)) \right]\!$}
    \\ \hline
    \end{tabular}
\vspace{-1mm}
\end{center}
\label{tbl:tbl1}
\end{table}

\section{Generalized master stability analysis for nonidentical oscillators}
\label{sec:msf}

The MSF formalism was originally introduced to study the stability of states of complete synchronization 
in networks of 
{\bb diffusively coupled} 
identical oscillators \cite{pecora1998master}.  Applied to the linearized dynamics around 
the synchronization manifold, it effectively reduces the dimension of the variational equation from the dimension $N\times d$
of the full system to the dimension $d$ of the local dynamics.
{\bb This is achieved}
 by simultaneously diagonalizing the Laplacian matrix and the
identity matrix, which can always be done when the Laplacian matrix is diagonalizable, as in the case of undirected networks. 
The approach has also been adapted to directed networks in which the Laplacian {\bb matrix} is not diagonalizable by replacing the diagonalization 
with a transformation into a Jordan canonical form \cite{nishikawa2006maximum}.

The main difficulty in extending this powerful formalism to the case of nonidentical oscillators is that the identity matrix is then
replaced by a set of more complicated matrices that, in general, cannot be simultaneously diagonalized with the 
{\bb coupling}
matrix.
{\bb We} address this problem by instead
{\bb finding the finest simultaneous block diagonalization of this set of matrices and the 
coupling
matrix,}
{\bb  corresponding to the largest possible dimension reduction from the original system that can be achieved by an orthogonal transformation matrix.}
Our approach is partially inspired by the framework of irreducible representation previously used to study cluster synchronization in networks of identical oscillators \cite{pecora2014cluster}, which 
{\bb leads to}
dimension reduction in that context.  
As shown below, our extension of the MSF formalism has the advantage of 
{\bb being applicable}
to both diffusive and non-diffusive coupling forms, and independently of whether the coupling matrix is diagonalizable or not, 
{\bb provided the conditions for complete synchronization are satisfied.}

\subsection{Networks with diffusive coupling}
\label{sec:ndc}

We consider a network of $N$ nonidentical dynamical units coupled diffusively as
\begin{equation}
	\dot{\bm{x}}_i = \bm{F}_i(\bm{x}_i) - \sigma \sum_{j=1}^{N} L_{i,j} \bm{H}(\bm{x}_j),
	\label{eq:7}
\end{equation}
where $\bm{x}_i$ is the $d$-dimensional state vector, $\bm{F}_i: \mathbb{R}^d \rightarrow \mathbb{R}^d$ is the vector field governing the uncoupled dynamics of the $i$-th oscillator, $\bm{L}$ the graph Laplacian encoding the (possibly weighted and directed) network structure, $\bm{H}$ is the interaction function, and $\sigma$ is the coupling strength. The Laplacian matrix $\bm{L}$ is defined as $L_{i,j} = \delta_{i,j}\mu_i - A_{i,j}$, where
$\delta_{i,j}$ is the Kronecker delta, $A_{i,j}$ is the entry of the adjacency matrix representing the connection from node $j$ to node $i$, and $\mu_i = \sum_j A_{i,j}$ is the 
{\bb (weighted)} indegree of node $i$. 
The vector field functions $\bm{F}_i$ are  chosen from a set of $M$ nonidentical functions $\{\bm{F}^{(\beta)}\}$. In a state of complete synchronization,  
{\bb the condition in equation~\eqref{eq:sync} holds}
for some 
{\bb orbit}
 $\bm{s}(t)$, which, together with equation~\eqref{eq:7}, implies 
 that 
$\dot{\bm{s}} (t)= \bm{F}_i(\bm{s}(t))$
for all $i$ and {\bb thus} that $\bm{F}_1(\bm{s}(t))=\bm{F}_2(\bm{s}(t)) =\cdots =\bm{F}_N(\bm{s}(t))$. Therefore, the necessary and sufficient condition for complete synchronization of diffusively coupled nonidentical oscillators is that all oscillators coincide on some common 
{\bb orbit}
$\bm{s}(t)$ that is a solution of each uncoupled oscillator {\bb (which in this case is equivalent to being a common orbit of the coupled oscillators).} Assuming this condition is satisfied, the question we address next is how  to establish an easily verifiable condition for the {\bb {\it stability}}  
of such complete 
{\bb synchronization} states.

The equation governing the evolution of a 
{\bb perturbation}
 from the 
 {\bb state}
 of complete synchronization can be obtained by linearizing 
equation~\eqref{eq:7}:
\begin{equation}
	\delta\dot{\bm{X}} = \left( \sum_{\beta=1}^M \bm{D}^{(\beta)} \otimes J\bm{F}^{(\beta)}(\bm{s}) - \sigma \bm{L} \otimes J\bm{H}(\bm{s}) \right) \delta\bm{X},
	\label{eq:8}
\end{equation}
where 
{\bb $\delta\bm{X} = \left(\delta \bm{x}_1^\intercal, \cdots, \delta \bm{x}_N^\intercal\right)^\intercal = \left((\bm{x}_1 - \bm{s})^\intercal, \cdots, (\bm{x}_N - \bm{s})^\intercal\right)^\intercal$ is the perturbation vector,} $\otimes$ denotes the Kronecker product, $J$ is the Jacobian operator, and $^\intercal$ indicates matrix transpose.
Denoting by $\mathcal{N}_\beta$ the set of nodes equipped with the $\beta$-th vector field function $\bm{F}^{(\beta)}$, the  $\bm{D}^{(\beta)}$ are $N\times N$ diagonal matrices given by 
\begin{equation}
	D_{i,i}^{(\beta)} =
	\begin{cases}
		1, & \quad \text{if } i \in \mathcal{N}_\beta, \\
		0, & \quad \text{otherwise}. \\
	\end{cases}
\end{equation}
Note that $\sum_{\beta=1}^M \bm{D}^{(\beta)} = \mathds{1}_N$, 
{\bb where} $\mathds{1}_N$ 
{\bb denotes}
the identity matrix.
It is generally impossible to find a basis of eigenvectors that {\bb would} simultaneously diagonalize the matrices $\{\bm{D}^{(\beta)}\}$ and $\bm{L}$.
Thus, to generalize the MSF formalism to the case of nonidentical oscillators
starting from 
equation \eqref{eq:8},  we must abandon the hope
of completely decoupling the perturbation modes in general. 
{\bb Indeed, the}
class of systems is too broad to allow a simple reduced form as in the original MSF formalism 
(this is the case also for previous generalizations applied to the problems of cluster synchronization \cite{pecora2014cluster} and nonidentical interaction 
functions \cite{Genioe1601679}). 
{\bb Informally,}  
because a transformation that simplifies some of these matrices would generally complicate others,
the key is to find the best balance between the competing goals of simplifying different matrices.

We now establish a formalism able to exploit partial decoupling 
among the perturbation modes
in equation \eqref{eq:8}. 
This is achieved through a transformation of equation~\eqref{eq:8} by 
a matrix $\bm{P}$ that implements 
{\bb the finest}
simultaneous block diagonalization (SBD) of all {\bb $\bm{D}^{(\beta)}$} and $\bm{L}$ (the notion of finest is defined rigorously below). 
This transformation will be referred 
to as the SBD transformation, and the corresponding coordinates as the SBD coordinates.
The transformation matrix $\bm{P}$ can be defined as an orthogonal matrix ({\bb generally} not unique) that decomposes the matrix $*$-algebra generated by $\{\bm{D}^{(1)}, \bm{D}^{(2)}, \cdots, \bm{D}^{(M)}, \bm{L} \}$ into 
{\bb the direct sum of (possibly multiple copies of)}
lower dimensional irreducible matrix $*$-algebras. 
For clarity, we defer to Sec.~\ref{sec:sbd}  the discussion on the construction of matrix $\bm{P}$ and proceed for the moment assuming that this matrix has been calculated.

The transformation matrix $\bm{P}$ applied to equation~\eqref{eq:8} leads to 
\begin{equation}
	\dot{\bm{\eta}} = \left( \sum_{\beta=1}^{M} \tilde{\bm{D}}^{(\beta)} \otimes J\bm{F}^{(\beta)}(\bm{s}) - \sigma \tilde{\bm{L}} \otimes J\bm{H}(\bm{s}) \right) \bm{\eta},
	\label{eq:9}
\end{equation}
where $\bm{\eta}=\left( \bm{P}^\intercal \otimes \mathds{1}_d \right) \delta\bm{X}$ is the 
{\bb perturbation vector}
expressed in the SBD coordinates. Here, the set of matrices $\{\tilde{\bm{D}}^{(\beta)}\} = \{\bm{P}^\intercal\bm{D}^{(\beta)}\bm{P}\}$ and $\tilde{\bm{L}} = \bm{P}^\intercal\bm{L}\bm{P}$ are block diagonal matrices with the {\it same} block structure. 
It is instructive to notice that the effect of the nonidentical Jacobians $\{J\bm{F}^{(\beta)}(\bm{s})\}$ on the variational equation~\eqref{eq:9} is analogous to that of the Jacobians $\{J\bm{F}(\bm{s}_m)\}$  associated with different synchronization states $\bm{s}_m$ of different clusters in the cluster synchronization 
{\bb systems}
studied in Ref.~\cite{sorrentino2016complete}. A key difference is that the assignment of $\{J\bm{F}(\bm{s}_m)\}$ in cluster synchronization is linked to $\bm{L}$ by symmetries of the network, whereas here $\{J\bm{F}^{(\beta)}(\bm{s})\}$ can be assigned arbitrarily
{\bb (through the choice of the matrices $\{\bm{D}^{(\beta)}\}$).} This implies a more flexible relation among the matrices in equation~\eqref{eq:9}, whose SBD transformation is in general not an irreducible representation transformation.

It is also instructive to notice that the original MSF formalism is recovered when $M = 1$ and the network is undirected 
{\bb (i.e., the Laplacian matrix is symmetric).}
In this case $\bm{D}^{(1)}=\tilde{\bm{D}}^{(1)}=\mathds{1}_N$, matrix $\bm{P}$ can be constructed from the eigenvectors of $\bm{L}$, and $\tilde{\bm{L}}$ is a diagonal matrix with the eigenvalues $v_i$ of $\bm{L}$ as diagonal elements. Letting $\bm{F} = \bm{F}^{(1)}$, equation~\eqref{eq:9} reduces to $N$ decoupled equations,
\begin{equation}
	\dot{\bm{\eta}}_i = \left( J\bm{F}(\bm{s}) - \sigma v_i J\bm{H}(\bm{s}) \right) \bm{\eta}_i,
	\label{eq:10}
\end{equation}
from which the MSF can be calculated as the maximal Lyapunov exponent (MLE) for different values of $v = \sigma v_i$.  

\begin{figure}[htb]
\centering
\subfloat[]{
\includegraphics[width=.3\columnwidth]{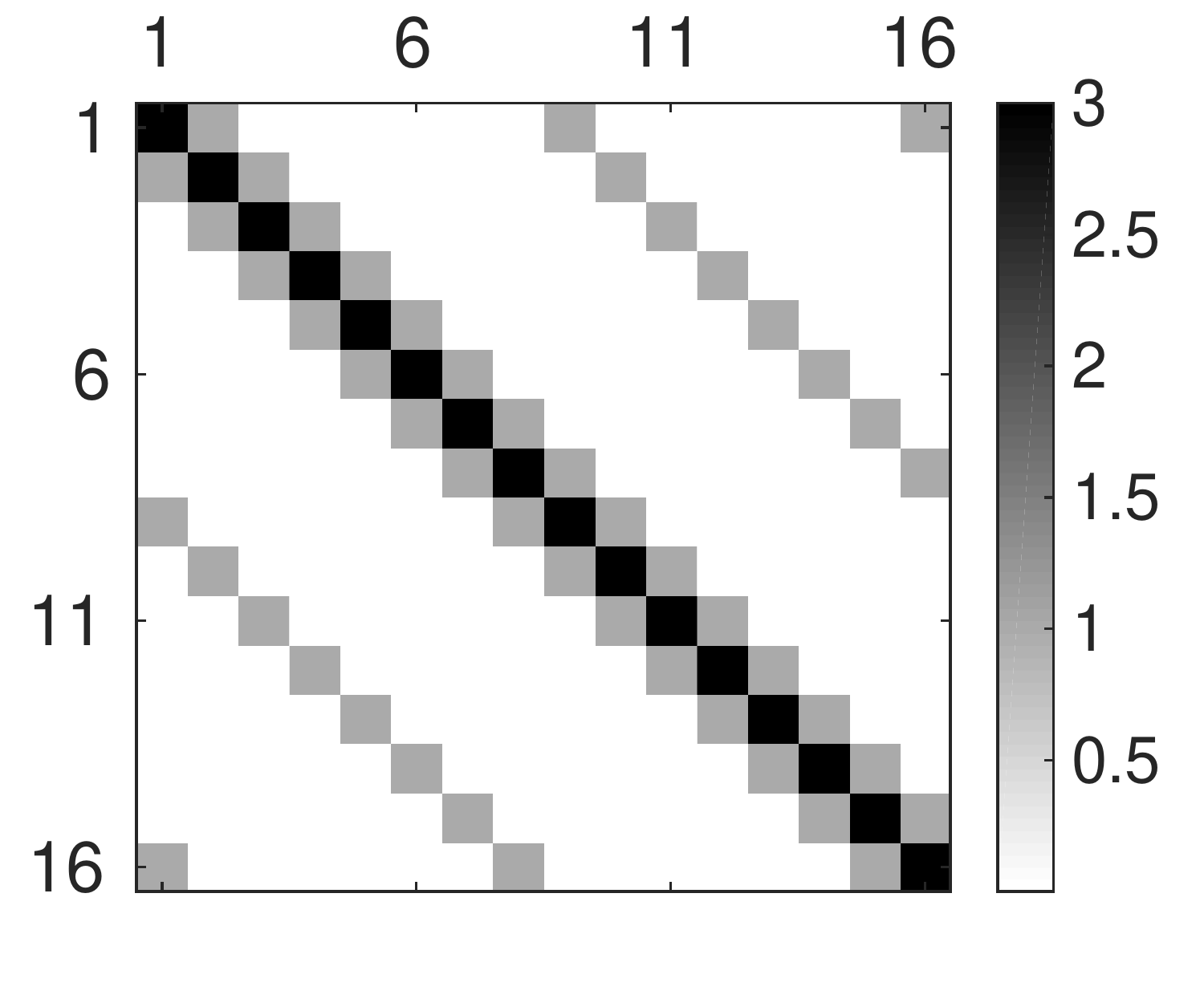}
}
\llap{\parbox[b]{2in}{(a)\\ \rule{0ex}{1.2in}}}
\hfil
\subfloat[]{
\includegraphics[width=.3\columnwidth]{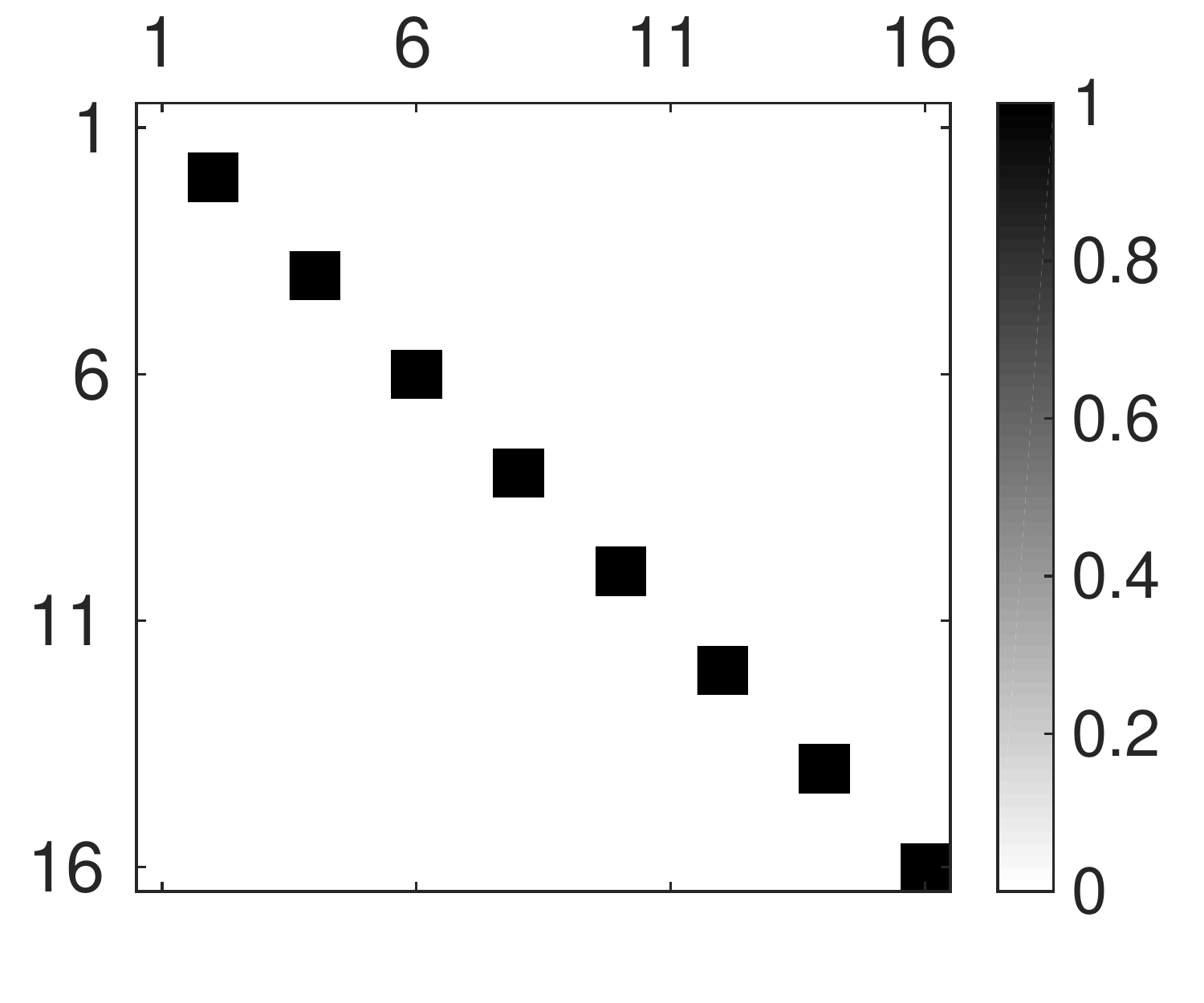}
}
\llap{\parbox[b]{2in}{(b)\\ \rule{0ex}{1.2in}}}
\subfloat[]{
\includegraphics[width=.3\columnwidth]{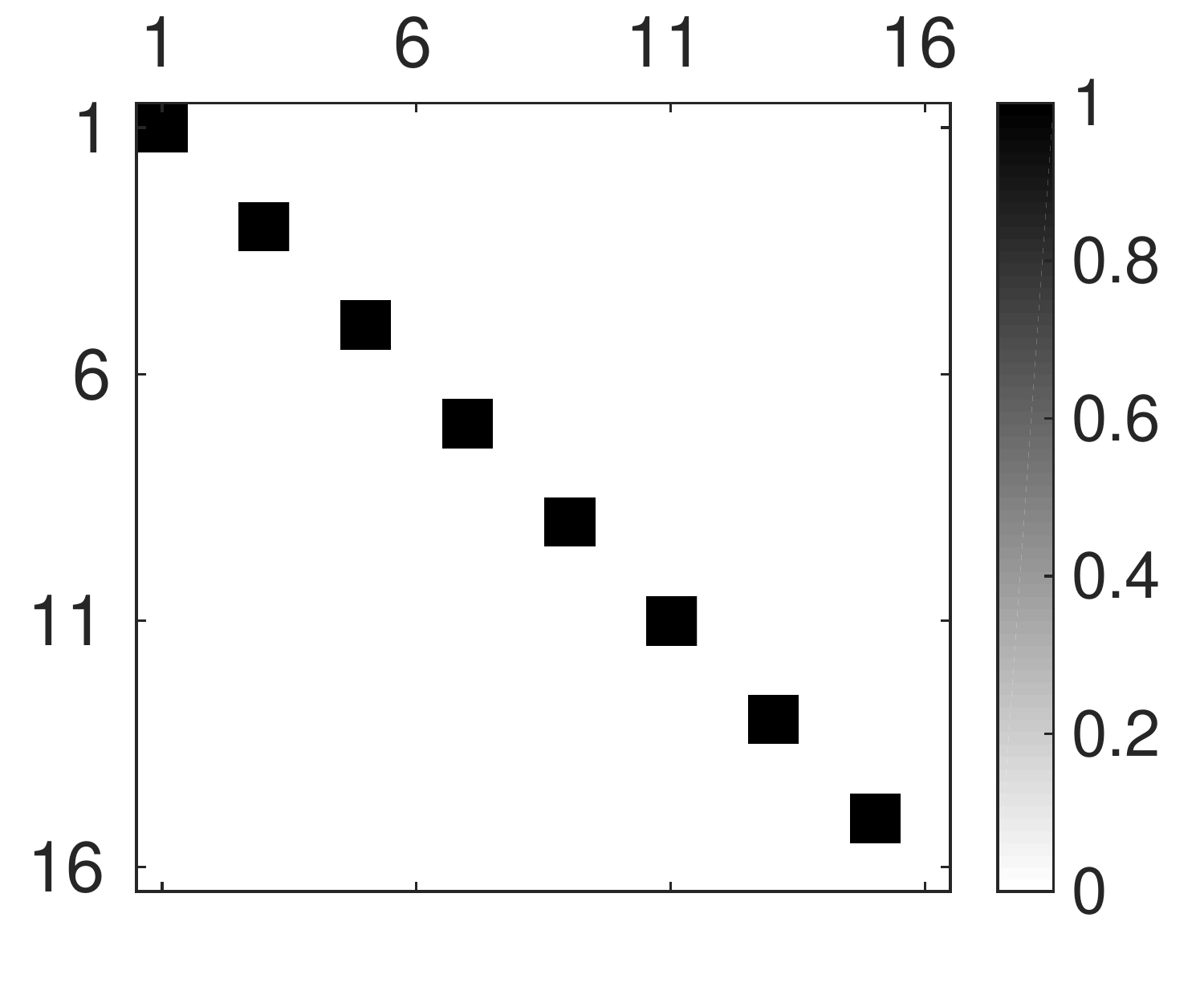}
}
\llap{\parbox[b]{2in}{(c)\\ \rule{0ex}{1.2in}}}
\\[-4ex]
\subfloat[]{
\includegraphics[width=.3\columnwidth]{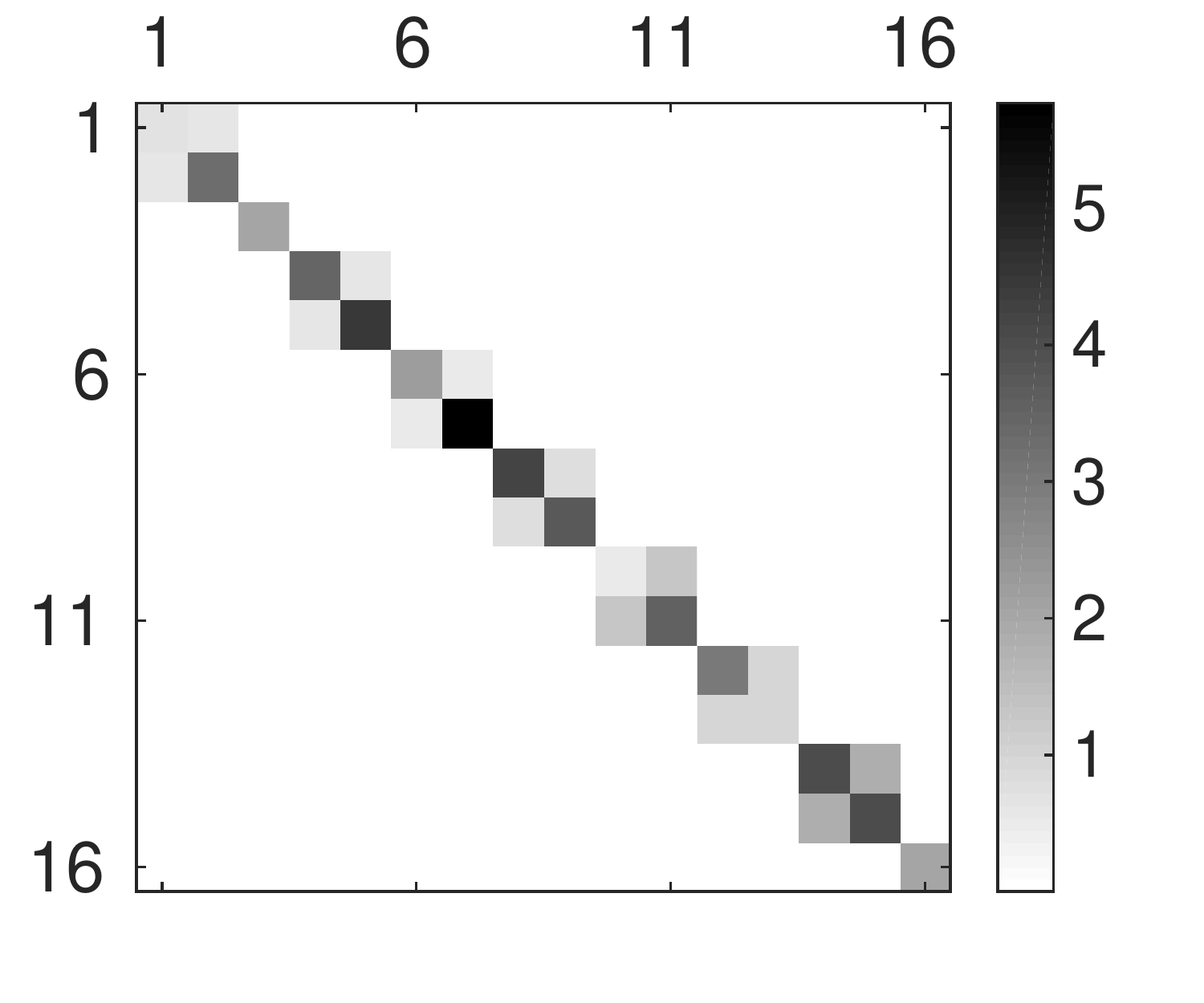}
}
\llap{\parbox[b]{2in}{(d)\\ \rule{0ex}{1.2in}}}
\hfil
\subfloat[]{
\includegraphics[width=.3\columnwidth]{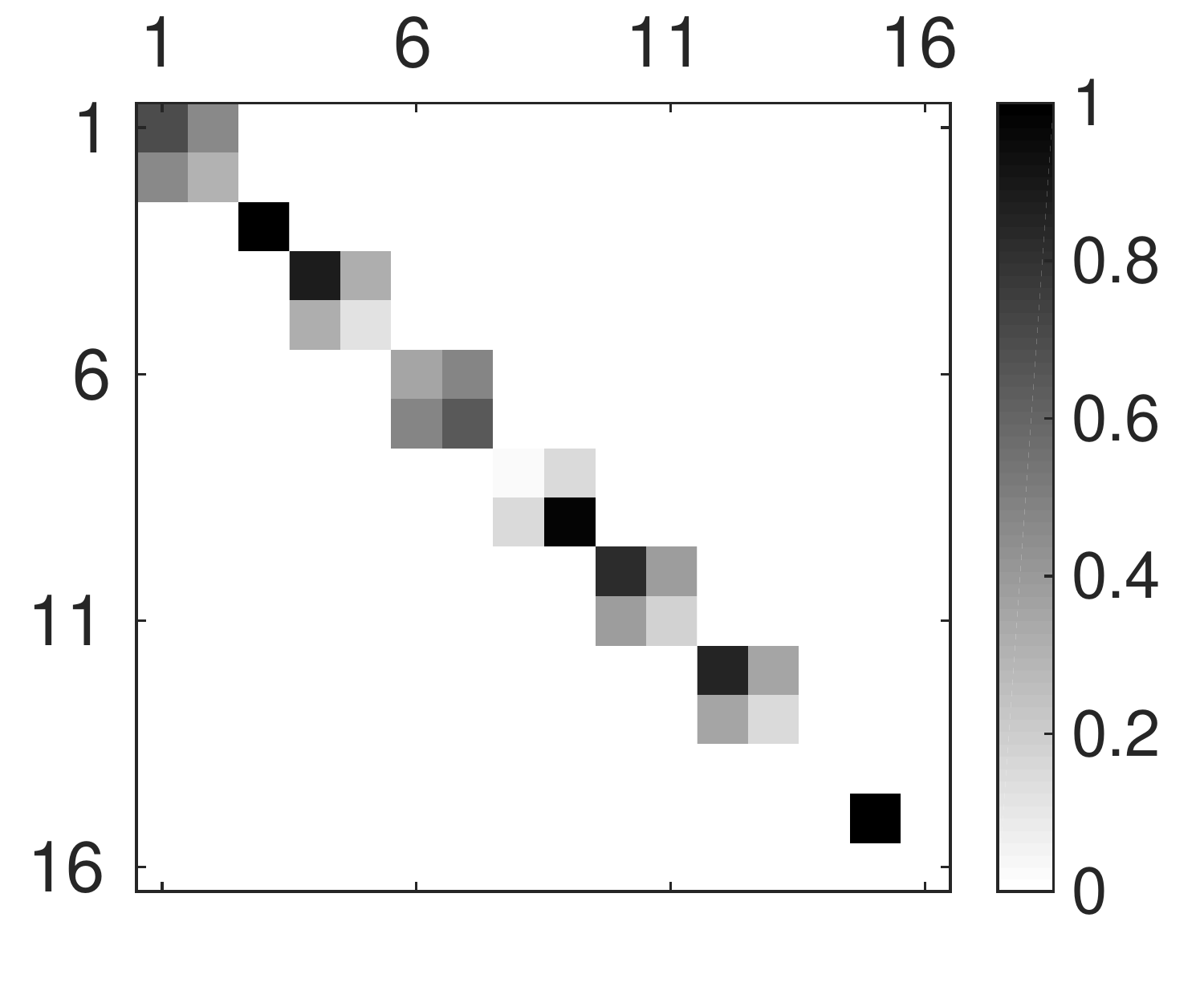}
}
\llap{\parbox[b]{2in}{(e)\\ \rule{0ex}{1.2in}}}
\hfil
\subfloat[]{
\includegraphics[width=.3\columnwidth]{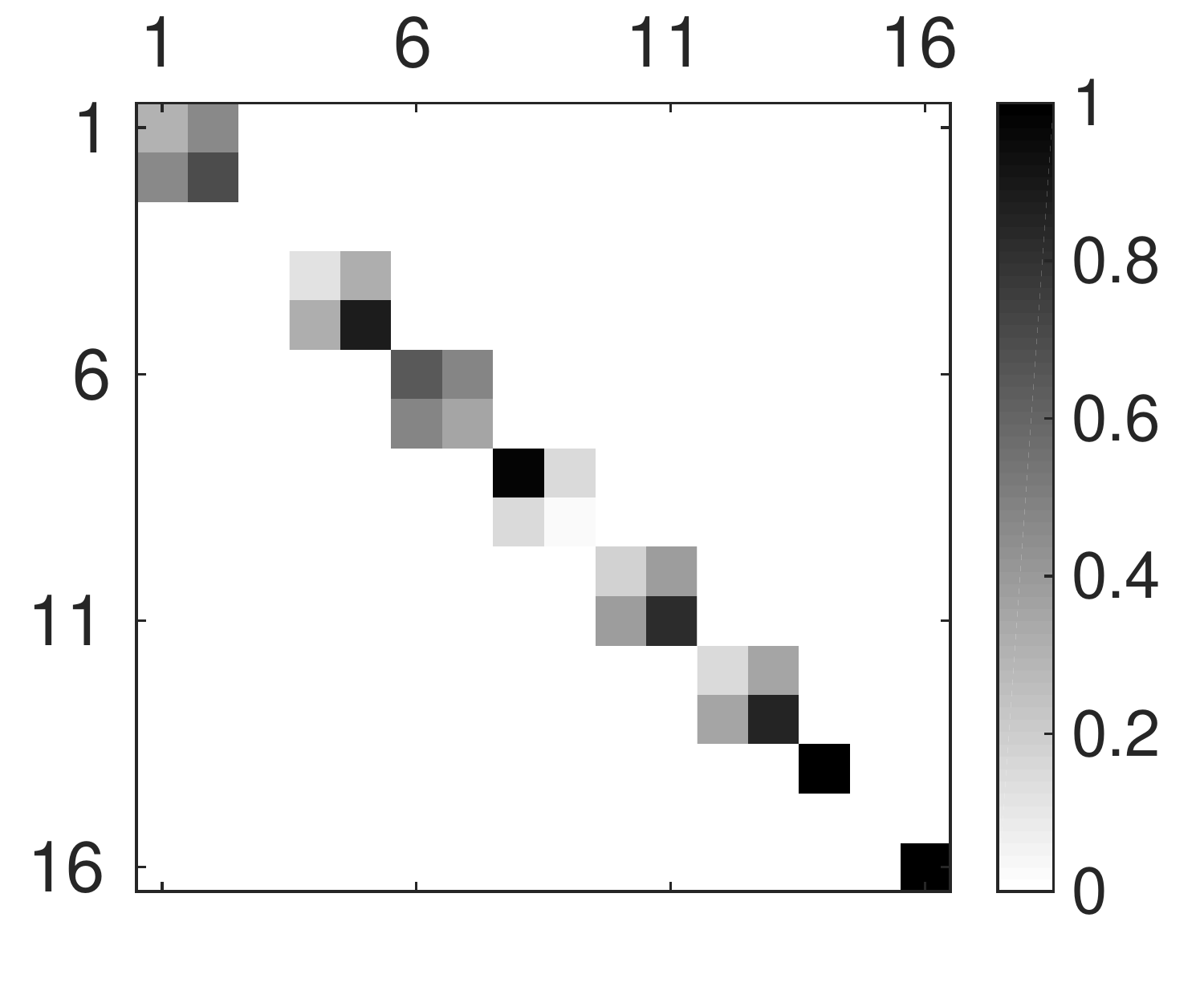}
}
\llap{\parbox[b]{2in}{(f)\\ \rule{0ex}{1.2in}}}
\vspace{-5mm}
\caption{(a) Matrix $\bm{L}$, (b) matrix $\bm{D}^{(1)}$, and (c) matrix $\bm{D}^{(2)}$  
representing an undirected wheel network with an alternating arrangement of two kinds of oscillators. 
{\bb (d-f) The corresponding matrices after the SBD transformation, 
showing that all transformed matrices share the same block structure.}
The grayscale indicates the absolute value of each element in the matrices.}
\label{fig:sbd}
\end{figure}

Numerical algorithms are available for the calculation of $\bm{P}$ given a set of matrices. 
Here, we adopt the method introduced in Ref.~\cite{maehara2011algorithm}, which we discuss in some detail in Sec.~\ref{sec:sbd}.
As an example, we simultaneously block diagonalize three $16 \times 16$ matrices: $\bm{L}$, $\bm{D}^{(1)}$, and $\bm{D}^{(2)}$. 
Matrix $\bm{L}$ is the Laplacian of an undirected 16-node wheel network (figure~\ref{fig:sbd}(a)), which is a ring network with additional connections between opposite nodes;  
 matrices $\bm{D}^{(1)}$ and $\bm{D}^{(2)}$ encode alternating arrangements of two kinds of oscillators {\bb (figure~\ref{fig:sbd}(b)-(c)).}
The matrices are simultaneously transformed into seven $2 \times 2$ blocks and two $1 \times 1$ blocks {\bb (figure~\ref{fig:sbd}(d)-(f)).}
Thus, the original equation of $16d$ dimensions can be reduced to seven equations of dimension $2d$ and two equations of dimension $d$, which significantly simplifies the stability analysis.

Equation~(\ref{eq:9}) is partially decoupled according to the block 
{\bb structure}
of $\{\tilde{\bm{D}}^{(\beta)}\}$ and $\tilde{\bm{L}}$.\ Therefore, we can calculate the Lyapunov exponents of the equations corresponding to each block 
{\bb separately. The block structure does not depend on the synchronization state $\bm{s}(t)$, but equation (\ref{eq:9}) itself does, and so do the associated Lyapunov exponents. 
One of the Lyapunov exponents corresponds to perturbations along the synchronization orbit $\bm{s}(t)$, and is zero whether this orbit is periodic or chaotic. 
The primary question of interest concerns the stability of the synchronization state $\bm{s}(t)$, which is determined by the maximal transverse Lyapunov exponent (MTLE). 
The MTLE always excludes the (null) Lyapunov exponent along $\bm{s}(t)$.
However, in contrast with the case of identical oscillators, more general perturbations of the form $\delta \bm{X} = \left(\delta\bm{x}^\intercal,\cdots,\delta\bm{x}^\intercal\right)^\intercal$ usually do not preserve synchronization and cannot be excluded upfront in the stability analysis in the case of non-identical oscillators.  
This is because the condition $\bm{F}_1(\bm{s}(t) + \delta \bm{x}_1(t)) = \cdots = \bm{F}_N(\bm{s}(t) + \delta \bm{x}_N(t))$ is generally not satisfied for all $t$ even when $\delta\bm{x}_1(0) = \cdots = \delta\bm{x}_N(0)$. 
Our approach takes this into account automatically since, unlike the MSF formalism, it does not discard the contribution from such perturbation modes in the calculation of the MTLE (instead it only excludes Lyapunov exponents associated with perturbations satisfying $\delta\bm{x}_1(t) = \cdots = \delta\bm{x}_N(t)$ for all $t$).

Systems in which the synchronization orbit $\bm{s}(t)$ is not unique may synchronize even when the individual orbits are unstable. This is the case when the distinct oscillators in the system share common dynamics in a neighborhood of a chaotic attractor, and thus share all the (uncountably many) orbits of the attractor as synchronization orbits. Due to chaos, those orbits are necessarily unstable, but parallel perturbations of the form
$\delta \bm{X} = \left(\delta\bm{x}^\intercal,\cdots,\delta\bm{x}^\intercal\right)^\intercal$ 
may merely change synchronization trajectory without destroying long-term synchronization.  
An example in which a full synchronization manifold $\bm{x}_1 = \cdots = \bm{x}_N$ is invariant for nonidentical oscillators is given in Sec.\ \ref{sec:nam}.


}


Finally, we note that our approach also applies to systems with nonidentical interaction functions by simultaneously block diagonalizing the set of Laplacian matrices that represent different types of interactions.
In particular, 
{\bb by finding the finest simultaneous block diagonalization of the set of all matrices in the $*$-algebra associated with the variational equation,}
our method 
{\bb can} provide a more significant dimension reduction than the coordinates proposed in Ref.~\cite{Genioe1601679} in the study of networks with multiple interaction layers, where partial diagonalization was implemented by choosing as a basis the eigenvectors of one of the Laplacian matrices from the set.
{\bb For identical oscillators, the utility of the SBD transformation in this context has been demonstrated in a separate study~\cite{irving2012synchronization}. 
Our 
formulation, however, applies to systems in which both the interactions and the oscillators are allowed to be nonidentical (provided they satisfy the conditions for complete synchronization).
Another advantage of 
{\bb the approach}
 is that, because it does not require the graph 
{\bb Laplacians} to be diagonalizable, its use extends to systems with directed couplings in general. 

\subsection{Networks with adjacency-matrix coupling}
\label{sec:nam}

The method developed in Sec.~\ref{sec:ndc} also applies to oscillator networks with adjacency-matrix coupling:
\begin{equation}
	\dot{\bm{x}}_i = \bm{F}_i(\bm{x}_i) + \sigma \sum_{j=1}^{N} A_{i,j} \bm{H}(\bm{x}_j),
	\label{eq:15}
\end{equation}
where $\bm{A}$ represents the adjacency matrix of the network and the other symbols are defined as in equation \eqref{eq:7}.  This form of coupling has been considered for the case of identical oscillators (i.e., identical $\bm{F}_i$  for all $i$) in the study of cluster (hence partial) synchronization  \cite{pecora2014cluster}.  In order to consider complete synchronization for nonidentical $ \bm{F}_i$, we first note that the {\bb necessary and sufficient} condition for the existence of a synchronous state 
{\bb $\bm{s}(t)$ as defined by equation~\eqref{eq:sync}} 
is that 
{\bb $\bm{F}_i(\bm{s}(t)) + \sigma \mu_i \bm{H}(\bm{s}(t))=\dot{\bm{s}}(t)$ 
holds for all $i$,}
where we recall that $\mu_i$ denotes the indegree of node $i$.

Following the same procedure used for diffusive coupling, we obtain an equation analogous to equation~\eqref{eq:9},
\begin{equation}
	\dot{\bm{\eta}} = \left( \sum_{\beta=1}^{M} \tilde{\bm{D}}^{(\beta)} \otimes J\bm{F}^{(\beta)}(\bm{s}) + \sigma \tilde{\bm{A}} \otimes J\bm{H}(\bm{s}) \right) \bm{\eta},
	\label{eq:16}
\end{equation}
where 
{\bb $\tilde{\bm{A}}=\bm{P}^\intercal \bm{A} \bm{P}$}
is the adjacency matrix $\bm{A}$ after the orthogonal transformation $\bm{P}$. As in the case of diffusive coupling,
this transformation reduces the dimension of the problem by 
partially decoupling the perturbation modes in the original equation.
{\bb
The stability of the synchronization orbit $\bm{s}(t)$ 
is now determined by 
considering the 
MTLE  associated with equation~\eqref{eq:16}, which always excludes the (null)  Lyapunov exponent associated 
with the perturbation mode along  this orbit (and it often excludes only this exponent, but see an exception below).}
%
%

As an example, consider $N$ nonidentical R\"ossler oscillators coupled through an undirected chain network. The equation for an isolated R\"ossler oscillator $\dot{\bm{x}}_i = \bm{F}_i(\bm{x}_i)$ is given by
\begin{equation}
	\begin{cases}
    \dot{x}_{i1} = -x_{i2} - x_{i3}, \\
    \dot{x}_{i2} = x_{i1} + a_i x_{i2},\\
    \dot{x}_{i3} = b_i + (x_{i1} - c_i) x_{i3}, 
	\end{cases}
\end{equation}
and the coupling function is taken to be $\bm{H}(\bm{x}_j) = (0, x_{j2}, 0)^\intercal$. For the two end nodes, $i=1$ and $N$,  we set the oscillator parameters to be $(a_i,b_i,c_i) = (0.1,0.2,9)$; for all the other nodes, $1 < i < N$, the parameters are  $(a_i,b_i,c_i) = (0,0.2,9)$. Since $\mu_i = 1$ for 
{\bb $i=1$ and $N$,} 
and $\mu_i = 2$ for all other $i$, there 
{\bb exist common orbits
$\bm{s}=(s_1,s_2,s_3)$}
such that $\bm{F}_i(\bm{s}(t)) + \sigma \mu_i \bm{H}(\bm{s}(t))$ do not depend on $i$ and 
{\bb are equal to}
 $\big(-s_2 - s_3, s_1 + 0.2 s_2, 0.2 + (s_1 - 9)s_3\big)^\intercal$ for the coupling strength $\sigma = 0.1$.
In this case,
{\bb the synchronization manifold is invariant and}
 $\bm{s}(t)$ is 
{\bb any orbit}
in the chaotic attractor of an isolated R\"ossler oscillator for parameters $(a,b,c) = (0.2,0.2,9)$; 
 {\bb thus, perturbations parallel to the synchronization manifold do not lead to desynchronization.}
(An {\it undirected} ring network of identical R\"ossler oscillators with parameters $(a_i,b_i,c_i) = (0,0.2,9)$ 
also 
{\bb admits} 
the same $\bm{s}(t)$ as 
complete 
{\bb synchronization states}
for the same $\bm{H}$ and $\sigma$, although 
{\bb the stability can be}
different in general.) 
{\bb For}
the chain of heterogeneous R\"ossler oscillators, we 
{\bb have}
\begin{equation}  
    J\bm{F}_i = 
     \begin{pmatrix}
      0 & -1  & -1 \\
      1 & a_i &  0 \\
      x_{i3} & 0 & x_{i1} - c_i
     \end{pmatrix}\quad\text{and}\quad
     J\bm{H} = 
     \begin{pmatrix}
      0 & 0 & 0 \\
      0 & 1 & 0 \\
      0 & 0 & 0 
     \end{pmatrix}.
     \label{eq:rossler1}
\end{equation}
For $N=4$
{\bb oscillators,}
{\bb equation~\eqref{eq:16} then}
leads to the following (non-unique) set of matrices with the same block structure:
\begin{equation}
\resizebox{.9\hsize}{!}{$
    \tilde{\bm{A}} = 
     \begin{pmatrix}
      0.60 &  0.24 &    0 &     0 \\
      0.24 & -1.60 &    0 &     0 \\
         0 &     0 & 1.60 &  0.22 \\
         0 &     0 & 0.22 & -0.60 \\
     \end{pmatrix},\quad
     \tilde{\bm{D}}^{(1)} = 
     \begin{pmatrix}
      0.81 & -0.39 &    0 &     0 \\
     -0.39 &  0.19 &    0 &     0 \\
         0 &     0 & 0.19 &  0.40 \\
         0 &     0 & 0.40 &  0.81 \\
     \end{pmatrix},\quad
     \tilde{\bm{D}}^{(2)} = 
     \begin{pmatrix}
      0.19 &  0.39 &     0 &     0 \\
      0.39 &  0.81 &     0 &     0 \\
         0 &     0 &  0.81 & -0.40 \\
         0 &     0 & -0.40 &  0.19 \\
     \end{pmatrix},
     $}
     \label{eq:rossler2}
\end{equation}
where we have two matrices $\tilde{\bm{D}}^{(\beta)}$ because this example has two types of oscillators.

A scenario of special interest in our application to {\it AISync} below is the one in which all $oscillators$ are identically coupled, which implies that 
{\bb
\begin{equation}
\mu_1=\mu_2=\cdots=\mu_N \equiv \mu,
\label{balance}
\end{equation} 
for some common indegree $\mu$.}
In this case, as in the case of diffusive coupling, the 
condition for the synchronous state to exist is thus that all 
{\bb $\bm{F}_i$}
coincide on some orbit $\bm{s}(t)$. 
However, in contrast with the case of diffusive coupling, in the case of adjacency-matrix coupling 
this orbit is generally not a solution of the uncoupled oscillator dynamics, 
{\bb but rather of $\dot{\bm{s}}(t)=\bm{F}_i(\bm{s}(t)) + \sigma \mu \bm{H}(\bm{s}(t))$.}

\subsection{Networks with delay coupling}
\label{sec:nwd}

An important generalization of equation~\eqref{eq:7} 
{\bb is}
 oscillator networks with delay coupling of the form
\begin{equation}
	\dot{\bm{x}}_i (t) = \bm{F}_i(\bm{x}_i(t)) + \sigma \sum_{j=1}^{N} A_{i,j} \left[\bm{H}(\bm{x}_j(t-\tau)) - \bm{H}(\bm{x}_i(t))\right],
	\label{eq:11}
\end{equation}
where $\tau$ is the time delay. 
{\bb Other forms of delay coupling are possible, including those that do not consider the self-feedback term $\bm{H}(\bm{x}_i(t))$ \cite{flunkert2010synchronizing,huber2003dynamics} and those that incorporate processing delay alongside the propagation delay $\tau$ \cite{yao2016effect,zou2013reviving}. Our framework applies to those scenarios as well, and here we focus on the coupling form in equation~\eqref{eq:11} for concreteness.}
We first note that, although for  $\tau=0$ this system reduces to the form of equation~\eqref{eq:7}, 
for  $\tau>0$ the coupling is no longer diffusive in the sense that the coupling term does not necessarily vanish in a state of complete synchronization.
{\bb In this case, the necessary and sufficient}
condition for the existence of a synchronous state 
{\bb $\bm{s}(t)$} 
 is that 
 {\bb $\bm{F}_i(\bm{s}(t)) + \sigma \mu_i \left[\bm{H}(\bm{s}(t-\tau)) - \bm{H}(\bm{s}(t))\right]=\dot{\bm{s}}(t)$ 
 holds for all $i$.}
Like 
{\bb in the case of}
adjacency-matrix coupling,  
{\bb this reduces to all $\bm{F}_i$ being equal along the orbit $\bm{s}(t)$
 when 
the condition in equation \eqref{balance} is satisfied,}
 which is the case 
 {\bb if}
  the 
oscillators are  identically coupled.

We can now extend the formalism established in Sec.~\ref{sec:ndc} to also include the delay-coupled system \eqref{eq:11}. As a sufficient condition for this extension, we will assume that the matrices $J\bm{F}^{(\beta)}(\bm{s})$ and $J\bm{H}(\bm{s})$ do not depend on time. 
{\bb Like}
in the other cases considered above, the oscillators  can in principle be arbitrarily different from each other as long as they coincide on the synchronization 
{\bb orbit}
 $\bm{s}(t)$, 
{\bb which is  generally not a solution of the isolated node dynamics.}
 No assumption need to be made about the structure of the network.

Using $\bm{\mu}$ to denote the diagonal matrix with $\mu_i$ in the $i$-th diagonal entry, the analog of equation~\eqref{eq:8} can be written as
\begin{equation}
	\delta\dot{\bm{X}}(t) = \left( \sum_{\beta=1}^M \bm{D}^{(\beta)} \otimes J\bm{F}^{(\beta)} - \sigma \bm{\mu} \otimes J\bm{H} \right) \delta\bm{X}(t) + \sigma \left( \bm{A} \otimes J\bm{H} \right) \delta\bm{X}(t-\tau).
	\label{eq:12}
\end{equation}
The 
SBD transformation can
{\bb then}
 be applied to $\{\bm{D}^{(\beta)}\}$, $\bm{\mu}$, and $\bm{A}$, to obtain
\begin{equation}
	\dot{\bm{\eta}} (t) = \left( \sum_{\beta=1}^M \tilde{\bm{D}}^{(\beta)} \otimes J\bm{F}^{(\beta)} - \sigma \tilde{\bm{\mu}} \otimes J\bm{H} \right) \bm{\eta}(t) + \sigma \left( \tilde{\bm{A}} \otimes J\bm{H} \right) \bm{\eta}(t-\tau),
	\label{eq:13}
\end{equation}
where $\tilde{\bm{\mu}}$ is the matrix $\bm{\mu}$ after the transformation.
If we now invoke the assumption that {\bb $\{J\bm{F}^{(\beta)}\}$} and $J\bm{H}$ are constant matrices {\bb on the orbit $\bm{s}(t)$,} it follows that the effect of the time delay $\tau$ in $\bm{\eta}(t - \tau)$ can be represented by the factor $e^{-\Lambda \tau}$ {\bb \cite{choe2010controlling},} 
resulting in the following transcendental characteristic equation for the exponent $\Lambda$:
\begin{equation}
	\det \left\{ \sum_{\beta=1}^M \tilde{\bm{D}}^{(\beta)} \otimes J\bm{F}^{(\beta)} + \sigma (e^{-\Lambda \tau} \tilde{\bm{A}} - \tilde{\bm{\mu}}) \otimes J\bm{H} - \Lambda\mathds{1}_{dN} \right\} = 0.
	\label{eq:14}
\end{equation}
Equation~\eqref{eq:14} can be factorized according to the common block structure of $\{\bm{D}^{(\beta)}\}$, $\bm{\mu}$, and $\bm{A}$. 
The Lyapunov exponents are then  obtained as $\text{Re}(\Lambda)$, where $\Lambda$ can be calculated efficiently for each block using already available root-finding algorithms.
The largest Lyapunov exponent calculated from the decoupled blocks corresponds to the MLE of the original full system.
{\bb The stability of a synchronization orbit $\bm{s}(t)$ is determined by the  MTLE associated with equation~\eqref{eq:13}, which, 
as in the previous cases, is determined by excluding the Lyapunov exponents associated with perturbations satisfying $\delta\bm{x}_1(t) = \cdots = \delta\bm{x}_N(t)$ for all $t$.}  


As an example, we note that the block diagonalized structure in figure~\ref{fig:sbd} also applies to equation~\eqref{eq:14} if we choose 
{\bb $\bm{A}$ to be the adjacency matrix of the same wheel network}
and the same 
arrangement of oscillators as represented by $\bm{D}^{(1)}$ and $\bm{D}^{(2)}$.  Since $\bm{\mu} = \tilde{\bm{\mu}} = \mu \mathds{1}_N$ in this case, the only difference between the sets of matrices to be block diagonalized 
{\bb in these two examples is that matrix $\bm{L}$ is now replaced by matrix $\bm{A}$.}
More generally, 
{\bb we can show}
that when the corresponding adjacency matrix $\bm{A}$ is in the matrix $*$-algebra generated by the Laplacian matrix $\bm{L}$ and $\{ \bm{D}^{(\beta)} \}$, the SBD transformation of $\bm{A}$ and $\{ \bm{D}^{(\beta)} \}$ always yields the same block structure as the one from $\bm{L}$ and $\{ \bm{D}^{(\beta)} \}$. This includes the case of identically coupled oscillators, where $\bm{A} = -\bm{L} + \mu \mathds{1}_N$.

\section{Finding the finest simultaneous block diagonalization}
\label{sec:sbd}

Having established in Sec.\ \ref{sec:msf} the usefulness of the SBD transformation in addressing the synchronization of nonidentical oscillators, 
we now consider this transformation 
rigorously. Moreover, we put 
{\bb on}
firm ground the notion of {\it finest simultaneous block diagonalization} 
and also discuss an algorithm for the calculation of the transformation matrix $\bm{P}$.

To define the SBD transformation 
 we must first introduce 
{\bb the}
 matrix 
 {\bb $*$-algebra~\cite{murota2010numerical},} 
 which is an object of study in non-commutative algebra. Denoting by  $\mathcal{M}_N$ the set of  $N \times N$ real matrices, a subset $\mathcal{T}$ of $\mathcal{M}_N$ is 
 a {\it matrix $*$-algebra} over $\mathbb{R}$ if 
\begin{equation}
	\bm{B},\,\bm{C} \in \mathcal{T};\,\alpha,\,\beta \in \mathbb{R} \implies \alpha\bm{B}+\beta\bm{C},\,\bm{BC},\,\bm{B}^\intercal
	\in \mathcal{T},
\end{equation}
{\bb and $\mathds{1}_N \in \mathcal{T}$.}
This structure is convenient because it is closed under the involution operation defined by the matrix transpose (thus the $*$).
{\bb We say a}
subspace $\mathcal{W}$ of $\mathbb{R}^N$ is 
$\mathcal{T}$-{\it invariant} if $\bm{B}\mathcal{W} \subseteq \mathcal{W}$ for every $\bm{B} \in \mathcal{T}$. 
 {\bb A matrix $*$-algebra  $\mathcal{T}$ is said to be {\it irreducible}
if $\{\bm{0}\}$ and $\mathbb{R}^N$  are the only $\mathcal{T}$-invariant subspaces.}
%

A matrix $*$-algebra $\mathcal{T}$ can always be {\bb decomposed, through an orthogonal matrix $\bm{P}$, 
into the direct sum of lower dimensional  matrix $*$-algebras that can be further decomposed into irreducible matrix $*$-algebras $\mathcal{T}_j$:
\begin{equation}
	\bm{P}^\intercal\mathcal{T}\bm{P} = \bigoplus_{j=1}^{\ell} \left( \mathds{1}_{m_j} \otimes \mathcal{T}_j \right)
	= \text{diag}\{\mathds{1}_{m_1} \otimes \mathcal{T}_1, \cdots, \mathds{1}_{m_\ell} \otimes \mathcal{T}_\ell\}.
	\label{eq:A-W}
\end{equation}
Here $\bigoplus$ denotes direct sum, $\ell$ is the number of irreducible matrix $*$-algebras in the decomposition, $m_j$ is the multiplicity of $\mathcal{T}_j$, 
and thus $\ell$ and/or $m_j$ are strictly larger than one unless $\mathcal{T}$ is already irreducible.}
{\bb The existence of such orthogonal matrix $\bm{P}$}
follows from Artin-Wedderburn  type structure theorems (Theorems~3.1 and 6.1 in Ref.~\cite{murota2010numerical}).
This decomposition implies that,  with a single orthogonal matrix $\bm{P}$, all 
matrices in $\mathcal{T}$ can be transformed simultaneously 
to a block diagonal form determined by equation~\eqref{eq:A-W}. 
{\bb The orthogonal matrix $\bm{P}$ in this equation is not unique, but the irreducible 
$*$-algebras
$\mathcal{T}_j$ are uniquely determined by $\mathcal{T}$ (up to isomorphism). 
That is, each diagonal block of the matrices after
{\bb an} SBD transformation is uniquely determined up to an orthogonal transformation.}

Now we are in a position to define precisely what we mean by the finest simultaneous block diagonalization of a given set of matrices. An orthogonal matrix $\bm{P}$ is said to give the {\it finest} simultaneous block diagonalization of a set of $N \times N$ real matrices $\mathcal{B} = \{\bm{B}_1,\cdots,\bm{B}_n\}$, if it leads to the irreducible decomposition of the matrix $*$-algebra generated by $\{\mathds{1}_N,\bm{B}_1,\cdots,\bm{B}_n\}$. 
It follows that the dimension of each diagonal block is finest also in the sense that it cannot be further reduced without violating the condition of it being a simultaneous block diagonalization for all matrices in the $*$-algebra, even if we allow non-orthogonal similarity transformation matrices. 

If we allow non-orthogonal transformation matrices, then there can be a stronger definition of finest simultaneous block diagonalization in the sense that,
after the transformation, the $i$-th common 
{\bb blocks}
of all matrices in $\mathcal{B}$ only share trivial invariant subspaces {\bb for all $i$.} 
{\bb However, in the important}
case in which 
{\bb all}
matrices in $\mathcal{B}$ are symmetric, 
{\bb such as the synchronization models of Sec.~\ref{sec:msf} when considered on undirected networks,}
this 
{\bb stronger definition}
is equivalent to 
{\bb the one above}
based on the irreducible decomposition of the matrix $*$-algebra. 
{\bb Thus,}
for symmetric matrices,  the orthogonal matrix $\bm{P}$ in equation~\eqref{eq:A-W} always gives the finest
simultaneous block diagonalization of $\mathcal{B}$ according to both definitions. Henceforth we shall refer 
to the finest simultaneous block diagonalization exclusively in the sense of matrix $*$-algebra.

We can now turn to the numerical calculation of the transformation matrix $\bm{P}$.
Algorithms for the determination of $\bm{P}$ given a set of matrices have been developed 
in previous studies  motivated by their applications in semidefinite programming and independent 
component analysis.  While to the best of our knowledge their potential for synchronization problems 
remains underexplored, we can in fact benefit quite directly from such algorithms in connection with 
the SBD transformations we consider. In this work we {\bb adopt}  
an implementation of the 
method introduced in Ref.~\cite{maehara2011algorithm}, which considers the commutant algebra of the matrix $*$-algebra 
generated by $\{\mathds{1}_N,\bm{B}_1,\cdots,\bm{B}_n\}$, defined as the set of matrices that commute with all matrices of that $*$-algebra;
this approach provides 
{\bb a} simpler algorithm than those working directly with the original matrix $*$-algebra \cite{murota2010numerical,maehara2010numerical}.
The algorithm finds $\bm{P}$ through numerical linear-algebraic computations 
{\bb (i.e.,} 
eigenvalue/eigenvector calculations) and does not require any algebraic 
structure to be known in advance. 
Using the notation $[\bm{B}_k,\,\bm{X}] = \bm{B}_k\bm{X} - \bm{X}\bm{B}_k$, we can  summarize the algorithm into two steps as follows.

\begin{algorithm} (Algorithm~3.5 in Ref.~\cite{maehara2011algorithm})
	\begin{itemize}
		\item Calculate a symmetric $N \times N$ matrix $\bm{X}$ as a generic solution of $[\bm{B}_k,\,\bm{X}] = \bm{0}$, $k = 1,\dots,n$.
		\item Calculate an orthogonal matrix $\bm{P}$ that diagonalizes matrix $\bm{X}$.
	\end{itemize}
\end{algorithm}

Here, a generic solution means a matrix $\bm{X}$ with no accidental eigenvalue degeneracy that is not enforced by $\{\bm{B}_k\}$. 
While the second step is straightforward using standard algorithms, 
the first step can be addressed by translating it into an eigenvector problem 
{\bb that can then be} solved efficiently using the Lanczos method.

The intuition behind this algorithm is that the common invariant subspaces among $\{\bm{B}_k\}$ can be captured by $\bm{X}$, and $\bm{P}$ automatically decomposes $\mathbb{R}^N$ into the direct sum of those invariant subspaces.  Note 
{\bb that,} even though $\bm{X}$ commutes with all $\bm{B}_k$ and $\bm{P}$ diagonalizes $\bm{X}$, the matrix $\bm{P}$ will generally not diagonalize the matrices $\bm{B}_k$ but rather block diagonalize them---this is true even for symmetric matrices. Moreover, since we do not limit ourselves to symmetric matrices, being simultaneously diagonalizable (or even diagonalizable at all) is not implied by having a null commutator.

\section{\textit{AISync} in delay-coupled  oscillator networks}
\label{sec:ais}

For the purpose of studying {\it AISync}, we require the network structure to be symmetric (i.e., all oscillators to be identically coupled), so that any system asymmetry can be attributed to oscillator heterogeneity. 
{\bb Formally,} 
a (possibly directed and weighted) network is said to be symmetric if all nodes belong to a single orbit under the action of the network's automorphism group, whose elements  can be represented by permutation matrices  that re-order the nodes while leaving the adjacency matrix invariant. This is a  generalization of the (undirected and unweighted) vertex-transitive graphs considered in algebraic graph theory, and makes precise the intuition 
{\bb that}
all nodes play the same structural role by requiring the existence of symmetry operations that map one node to any other node {\bb  in the network.\footnote{\bb 
A network being symmetric should not be confused with a network having a symmetric coupling matrix,
which is neither sufficient nor necessary for the network to be symmetric.} }

Network symmetry implies 
the {\bb condition in equation \eqref{balance}.}  
Thus, 
{\bb for symmetric networks,} 
the condition for complete synchronization of nonidentical oscillators in the cases of adjacency-matrix coupling~\eqref{eq:15} and delay coupling~\eqref{eq:11} is that the vector field 
{\bb functions satisfy} 
$\bm{F}_1(\bm{s}) = \bm{F}_2(\bm{s}) =\cdots = \bm{F}_N(\bm{s})$ 
{\bb (as in the case of diffusive coupling~\eqref{eq:7}),}  where the synchronous state $\bm{s}=\bm{s}(t)$ is now a common solution of the coupled dynamics of all oscillators. Because the oscillators are nonidentical,  these equalities generally do not hold for 
{\bb state-space}
points outside the orbit {\bb $\bm{s}(t)$,} 
which 
{\bb can impact}
the stability of 
{\bb this orbit}
as a synchronous state solution.

Given $\{\bm{F}^{(\beta)}\}$, we say a system 
exhibits {\it AISync} if it satisfies the following two conditions: 
{\bb 1)} there are no stable states of complete synchronization for any homogeneous system (i.e.,   
any system for which $\bm{F}_1(\bm{x}) = \bm{F}_2(\bm{x}) = \cdots = \bm{F}_N(\bm{x}) $ {\bb $\forall \bm{x}$);}   
{\bb 2)} there is a heterogeneous system (i.e., 
a system such that  $\bm{F}_i \neq \bm{F}_j$ for some $i \neq j$) for which a stable synchronous state exists. Using the formalism presented above, now we show that {\it AISync} 
{\bb occurs}
in networks of delay-coupled Stuart-Landau oscillators. We also show that the scenario in which stable synchronization requires oscillators to be nonidentical extends naturally to non-symmetric 
{\bb networks.}

\subsection{Stuart-Landau oscillators sharing a common orbit}
\label{sec:ind}

We start with $N$ delay-coupled identical (supercritical) Stuart-Landau oscillators, whose equation in complex variable notation reads
\begin{equation}
	\dot{z}_j(t) = f(z_j(t)) + \sigma \sum_{k=1}^N A_{j,k} \left[ z_k(t-\tau) - z_j(t) \right],
	\label{eq:1}
\end{equation}
where $z_j=r_je^{i\psi_j}\in \mathds{C}$ for $\tau$ and $\sigma$ as in equation~\eqref{eq:11}.
The adjacency matrix $\bm{A}$ represents the structure of a symmetric network  and 
thus has a common row sum $\mu = \sum_k A_{j,k}$ $\forall j$.
Because the common row sum condition
{\bb is equivalent to the condition in equation \eqref{balance} and is thus} 
satisfied by any network 
with the same indegree for all nodes $j$, our analysis also applies to arbitrary non-symmetric network structures
that satisfy this indegree condition (including all directed regular graphs), as illustrated {\bb below} in Sec.~\ref{sec:nonsym}.

The local dynamics of each oscillator is given by the normal form of a supercritical Hopf bifurcation \cite{kuramoto2012chemical}:
\begin{equation}
	f(z_j) = \left[ \lambda + i\omega - (1+i \gamma)|z_j|^2 \right]z_j,
	\label{eq:2}
\end{equation}
where $\lambda$, $\omega$, and $\gamma$ are real parameters.  Intuitively, $\lambda$ relates to the amplitude of the oscillation, $\omega$ represents the base angular velocity, and $\gamma$ controls the amplitude-dependent angular velocity term.

Substituting  the limit cycle ansatz $z_j = r_0 e^{i \Omega t}$ into equation~\eqref{eq:1} and assuming  $r_0^2 \neq 0$, we obtain the invariant solution 
\begin{subequations}
    \begin{align}
      r_0^2 &= \lambda + \mu \sigma (\cos \Phi - 1), \\
      \Omega &=\omega - \gamma r_0^2 + \mu \sigma \sin \Phi,
    \end{align}
    \label{eq:3}
\end{subequations}
where we use the notation  $\Phi = - \Omega \tau$. 
This set of equations can be solved for $\Omega$ by substituting the $r_0^2$ in equation~(\ref{eq:3}b) with the right hand side of equation~(\ref{eq:3}a) and solving numerically the resulting transcendental equation. After determining $\Omega$, the value of $r_0^2$ can be immediately calculated from equation~(\ref{eq:3}a). 
There can be multiple solutions of $\Omega$ (thus also of $r_0^2$) for certain combinations of parameters (including spurious solutions with $r_0^2 < 0$).  Here, we focus on regions where a unique solution of positive $r_0^2$ exists. In addition, there {\bb is} always  
a time-independent solution $r_0=0$, corresponding to an amplitude death state, which was excluded in our derivation of equation~\eqref{eq:3} but can be identified directly from equations~\eqref{eq:1} and \eqref{eq:2}.

{\bb For identical Stuart-Landau oscillators,}
a variational equation for the limit cycle 
{\bb synchronous}
state ($z_j = r_0 e^{i \Omega t}$) 
is derived 
{\bb in Ref.~\cite{choe2010controlling}}
as
\begin{equation}
	\dot{\bm{\xi}}(t) = \mathds{1}_N \otimes (\bm{J}_0 - \mu \sigma \bm{R}) \bm{\xi}(t) + \sigma(\bm{A} \otimes \bm{R}) \bm{\xi}(t - \tau),
	\label{eq:4}
\end{equation}
where  
$\bm{J}_0 = \bigl(\begin{smallmatrix}
-2r_0^2 & ~0 \\ -2 \gamma r_0^2 & ~0
\end{smallmatrix} \bigr)$
and
$\bm{R} = \bigl(\begin{smallmatrix}
\cos \Phi & -\sin \Phi \\ \sin \Phi & ~~\cos \Phi
\end{smallmatrix} \bigr)$.
The $2N$-dimensional perturbation vector
is defined as 
{\bb $\bm{\xi} = \left(\bm{\xi}_1^\intercal,\cdots,\bm{\xi}_N^\intercal\right)^\intercal$,}
where $\bm{\xi}_j = (\delta r_j, \delta \psi_j)^\intercal$ and $(r_j,\psi_j) = \big(r_0(1 + \delta r_j),\Omega t + \delta \psi_j\big)$.
Equation~\eqref{eq:4} is a special case of equation~\eqref{eq:12} obtained by setting $M = 1$, $J\bm{F}^{(1)} = \bm{J}_0$, and $J\bm{H} = \bm{R}$.
Because the oscillators are identical, one can apply the standard MSF formalism to diagonalize $\bm{A}$ and obtain decoupled variational equations of the form
\begin{equation}
	\dot{\bm{\eta}}_k(t) = \bm{J}_0 \bm{\eta}_k(t) - \sigma \bm{R}\left[ \mu \bm{\eta}_k(t) - v_k \bm{\eta}_k(t - \tau) \right],
	\label{eq:5}
\end{equation}
with $\bm{\eta}_k$ representing the  perturbation vector associated with the eigenvalue $v_k$ of $\bm{A}$ after diagonalization. In particular, $v_0 = \mu$ corresponds to the perturbation mode (eigenvector) along the 
synchronization manifold.  Since $\bm{J}_0$ and $\bm{R}$ are constant matrices, the Lyapunov exponents of the perturbation modes are 
obtained as $\text{Re} (\Lambda)$, where the exponents  $\Lambda$  can be determined from the characteristic equation
\begin{equation}
	\det\{ \bm{J}_0 - \Lambda \mathds{1}_2 + (- \sigma \mu + \sigma v_k e^{-\Lambda \tau}) \bm{R} \} = 0.
	\label{eq:6}
\end{equation}
As usual, the resulting MLE can be interpreted as the MSF and visualized on the complex plane parametrized by the effective coupling parameter~$v = \sigma v_k$.

\begin{figure}[htb]
\centering
\subfloat[]{
\begin{tikzpicture}
    \node[anchor=south west,inner sep=0] (image) at (0,0,0) {\includegraphics[width=.45\columnwidth]{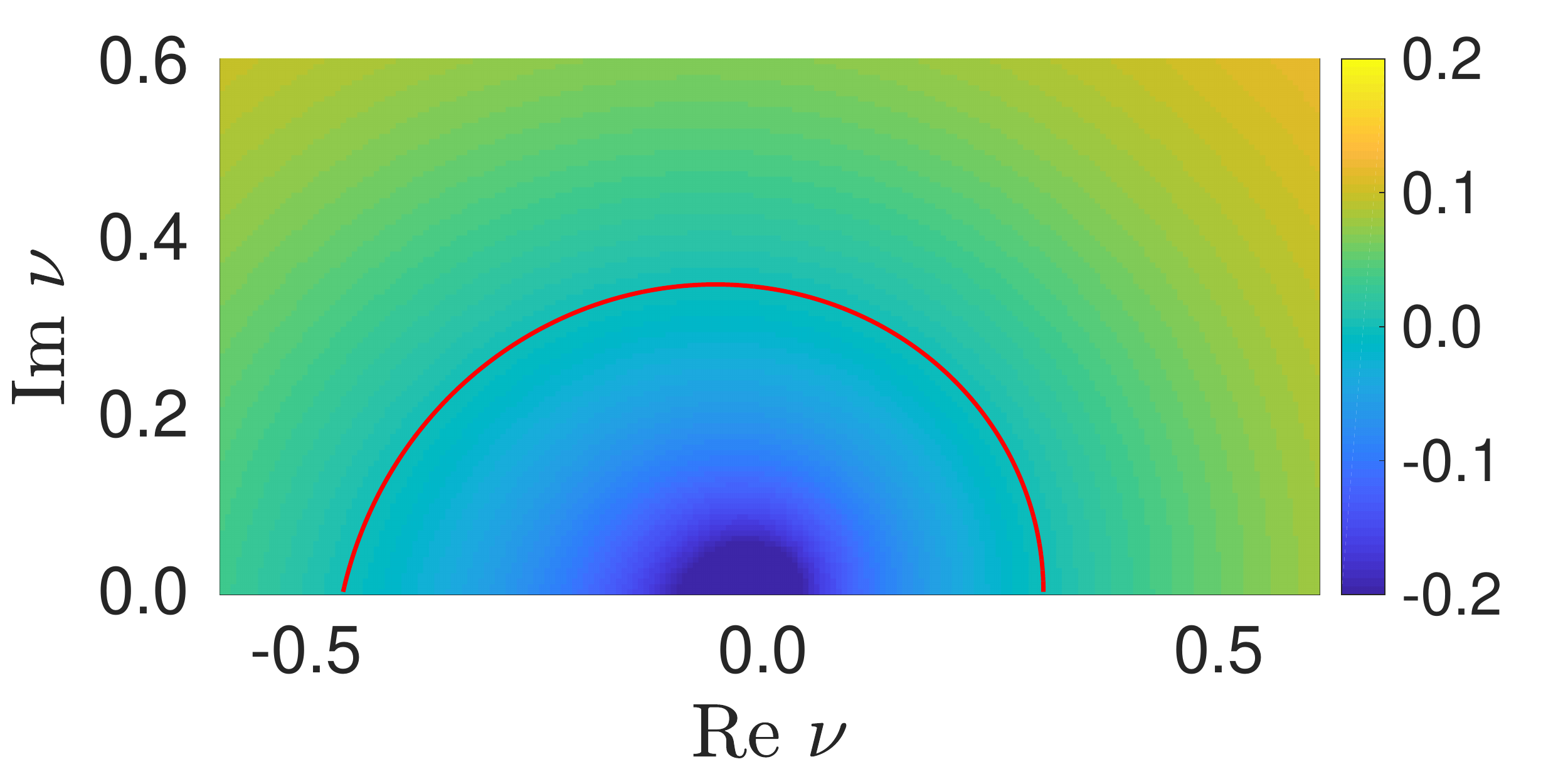}};
    \begin{scope}[x={(image.south east)},y={(image.north west)}]
    \node[rotate = 90] at (1,.55) {\scriptsize MLE};
    \end{scope}
\end{tikzpicture}
}
\llap{\parbox[b]{3in}{(a)\\\rule{0ex}{1.2in}}}
\hfil
\subfloat[]{
\begin{tikzpicture}
    \node[anchor=south west,inner sep=0] (image) at (0,0,0) {\includegraphics[width=.45\columnwidth]{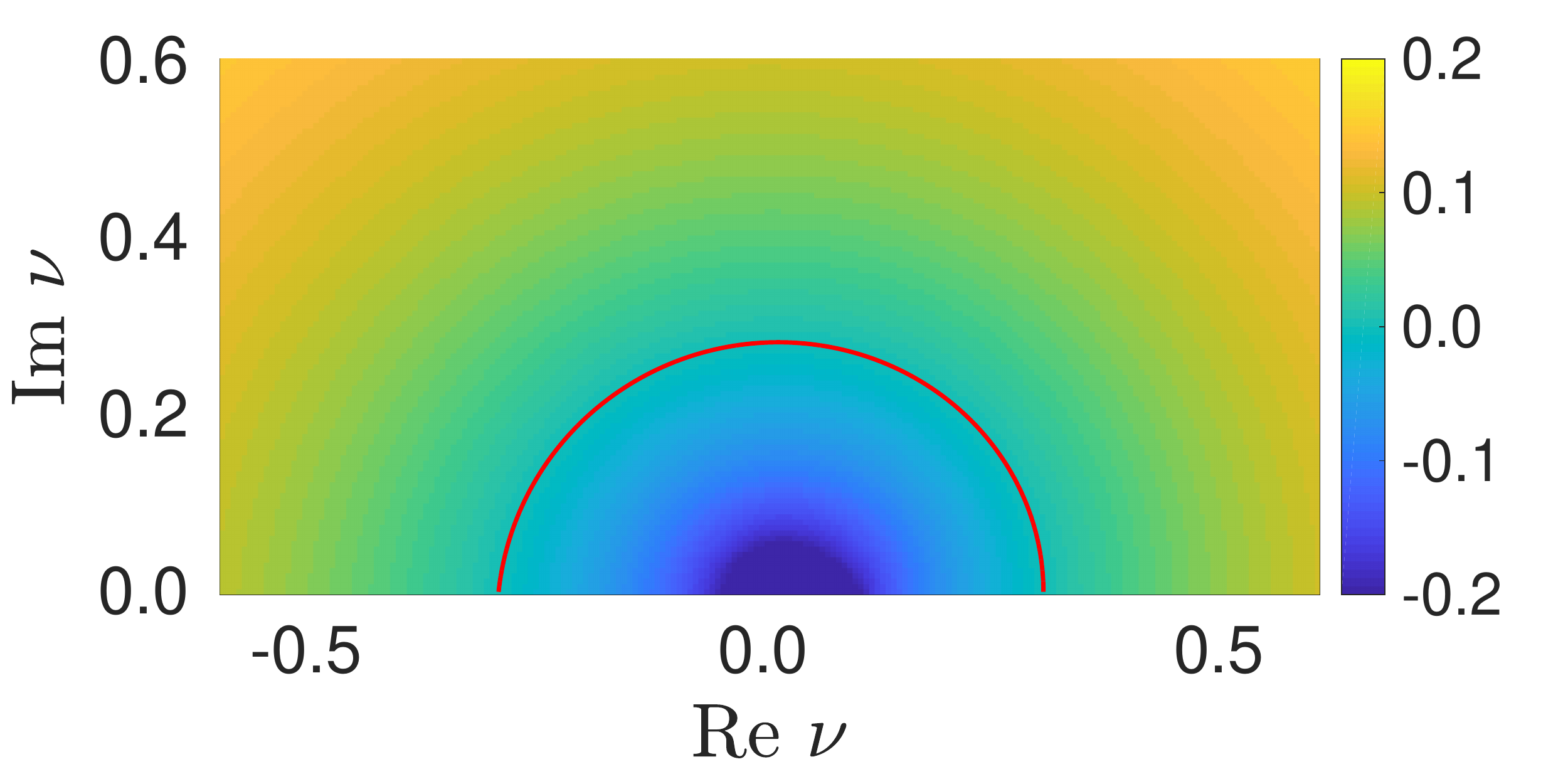}};
    \begin{scope}[x={(image.south east)},y={(image.north west)}]
    \node[rotate = 90] at (1,.55) {\scriptsize MLE};
    \end{scope}
\end{tikzpicture}
}
\llap{\parbox[b]{3in}{(b)\\\rule{0ex}{1.2in}}}
\vspace{-8mm}
\caption{MSF calculated from equation~\eqref{eq:6} for (a)  $h = 0$ and 
{\bb (b) $h = 0.8$.} 
The red contours mark the boundary of linear stability (i.e., where the MLE changes sign).
 The other parameters are $\sigma \mu = 0.3$, $\tau = 1.8\pi$, $\lambda_0 = 0.1$, $\omega_0 = 1$, and $\gamma_0 = 0$, which are the values used throughout the rest of the 
 {\bb article,}
  except when indicated otherwise.
}
\label{fig:msf}
\end{figure}

For a given Stuart-Landau oscillator $f(z;\lambda_0,\omega_0,\gamma_0)$ along with fixed parameters $\sigma \mu$ and $\tau$, there exists an entire class of nonidentical Stuart-Landau oscillators $\mathcal{SL}(\lambda_0,\omega_0,\gamma_0)$ characterized by a parameter $h$, consisting of oscillators of the form $f(z;\lambda_0,\omega_0 + h,\gamma_0 + h/r_0^2)$ for all values of $h\in \mathbb{R}$. All oscillators in this class share the common orbit $s = r_0 e^{i \Omega t}$ according to equation~\eqref{eq:3}, and are thus potential candidates for {\it AISync}. 
Moreover, 
 $\gamma = \gamma_0 + h/r_0^2$
enters the variational equation~\eqref{eq:4} through $\bm{J}_0$, so in general one would expect different stability of the synchronous solution for oscillators in $\mathcal{SL}(\lambda_0,\omega_0,\gamma_0)$ with different $h$
if the delay $\tau$ is nonzero (if $\tau = 0$, then $\bm{R}$ becomes diagonal and the off diagonal term $-2 \gamma r_0^2$ 
{\bb in $\bm{J}_0$}
will not contribute to the characteristic equation). This dependence on $h$ is illustrated in the example of figure~\ref{fig:msf}. Note that the amplitude death state, $z_j = 0$ for all $j$, is also {\bb a solution} common to all oscillators in $\mathcal{SL}(\lambda_0,\omega_0,\gamma_0)$.

\subsection{{\bb Demonstration} 
of \textit{AISync} in delay-coupled Stuart-Landau oscillators}
\label{sec:nind}

We now apply the framework developed thus far to characterize the {\it AISync} property  in  networks  of identically coupled Stuart-Landau oscillators. As a concrete example, we consider a directed ring network of $N=6$ nodes populated with  Stuart-Landau oscillators of two kinds:
$\bm{F}^{(1)}(z) = f(z;\lambda_0,\omega_0 + h,\gamma_0 + h/r_0^2)$ and $\bm{F}^{(2)}(z) = f(z;\lambda_0,\omega_0 - h,\gamma_0 - h/r_0^2)$ in the notation of Sec.~\ref{sec:msf},  
{\bb for}
$\lambda_0 = 0.1$, $\omega_0 = 1$, $\gamma_0 = 0$, and $h$ (which we convention to be positive from this point on) 
{\bb serving as a measure of}
the heterogeneity among oscillators. 
The other parameters are set to be $\sigma \mu = 0.3$ and $\tau = 1.8\pi$. Both $\bm{F}^{(1)}$ and $\bm{F}^{(2)}$ belong to $\mathcal{SL}(0.1,1,0)$ and thus satisfy $\bm{F}^{(1)}(s) = \bm{F}^{(2)}(s)$. Each of the six nodes in the directed ring network can be chosen as $\bm{F}^{(1)}$ or $\bm{F}^{(2)}$, which results in 
{\bb two}
possible homogeneous systems---$\bm{F}^{(1)}$ in all nodes (referred to as $+h$) or $\bm{F}^{(2)}$ in all nodes (referred to as $-h$)---and 11 distinct heterogeneous systems.

Equation~\eqref{eq:14} with $\bm{\mu} = \mu \mathds{1}_N$ can be applied to any of the 13 systems above. The block structure of $\tilde{\bm{D}}^{(1)}$, $\tilde{\bm{D}}^{(2)}$ and $\tilde{\bm{A}}$ varies from system to system. For example, when $\bm{F}^{(1)}$ and $\bm{F}^{(2)}$ are arranged on the ring such that every other oscillator is identical (corresponding to $\bm{D}^{(1)} = \text{diag}\{1,0,1,0,1,0\}$ and $\bm{D}^{(2)} = \text{diag}\{0,1,0,1,0,1\}$), they share a common block structure of one $2 \times 2$ block and one $4 \times 4$ block. This reduction from an $N \times N$ system of equations may seem small for $N=6$, but for any directed ring network of an even number of nodes $N$, the largest block will always be $4 \times 4$, which represents a tremendous simplification from the fully coupled set of equations when $N$ is large. 

Figure~\ref{fig:mtle} shows results of applying this method to calculate the MTLE of all heterogeneous systems (gray and blue lines) and {\bb of the} corresponding homogeneous systems $+h$ (cyan line) and $-h$ (red line). Blue marks the most synchronizable heterogeneous system, which has $\bm{D}^{(1)} = \text{diag}\{1,1,1,0,1,0\}$ and $\bm{D}^{(2)} = \text{diag}\{0,0,0,1,0,1\}$, and is referred to as $\pm h$. The limit cycle synchronous state of system $\pm h$ is stable for the entire range of $h$ considered.  For the homogeneous system $-h$, the synchronous state is stable for $h \in (0,0.23)$, and for the homogeneous system $+h$ this state is stable for $h \in (0,0.32)$. This gives rise to a wide {\it AISync} region, ranging from $h = 0.32$ all the way to at least $h = 1.0$ (the largest value considered in our calculations).  We also verified that the other possible symmetric state---the amplitude death state---is unstable for both homogeneous systems, and we show the corresponding MLEs in the inset of figure~\ref{fig:mtle}.

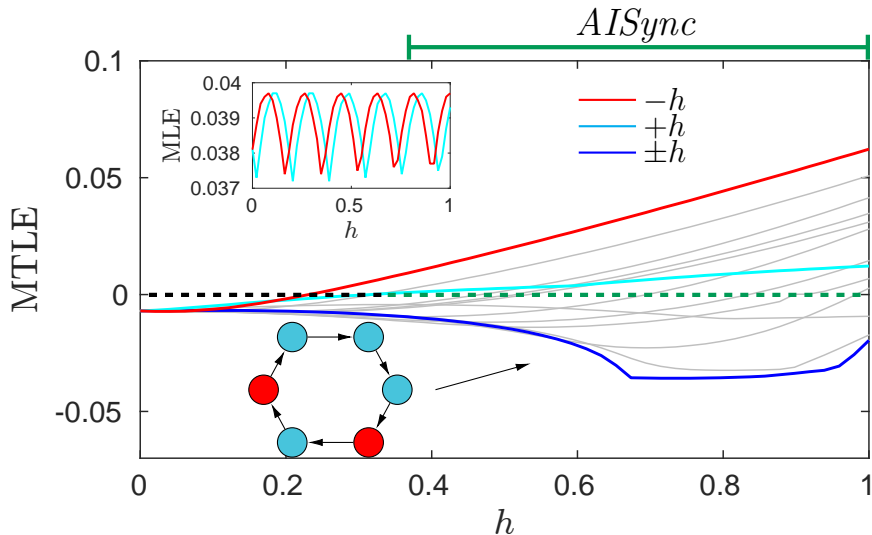
\begin{figure}[htb]
\centering
\subfloat[]{
\begin{tikzpicture} 
    [vertex/.style={draw,circle,thin},
    arc/.style={draw,-{Latex[length=2mm, width=1mm]}}]
    \node[anchor=south west,inner sep=0] (image) at (0,0,0) {\includegraphics[width=.8\columnwidth]{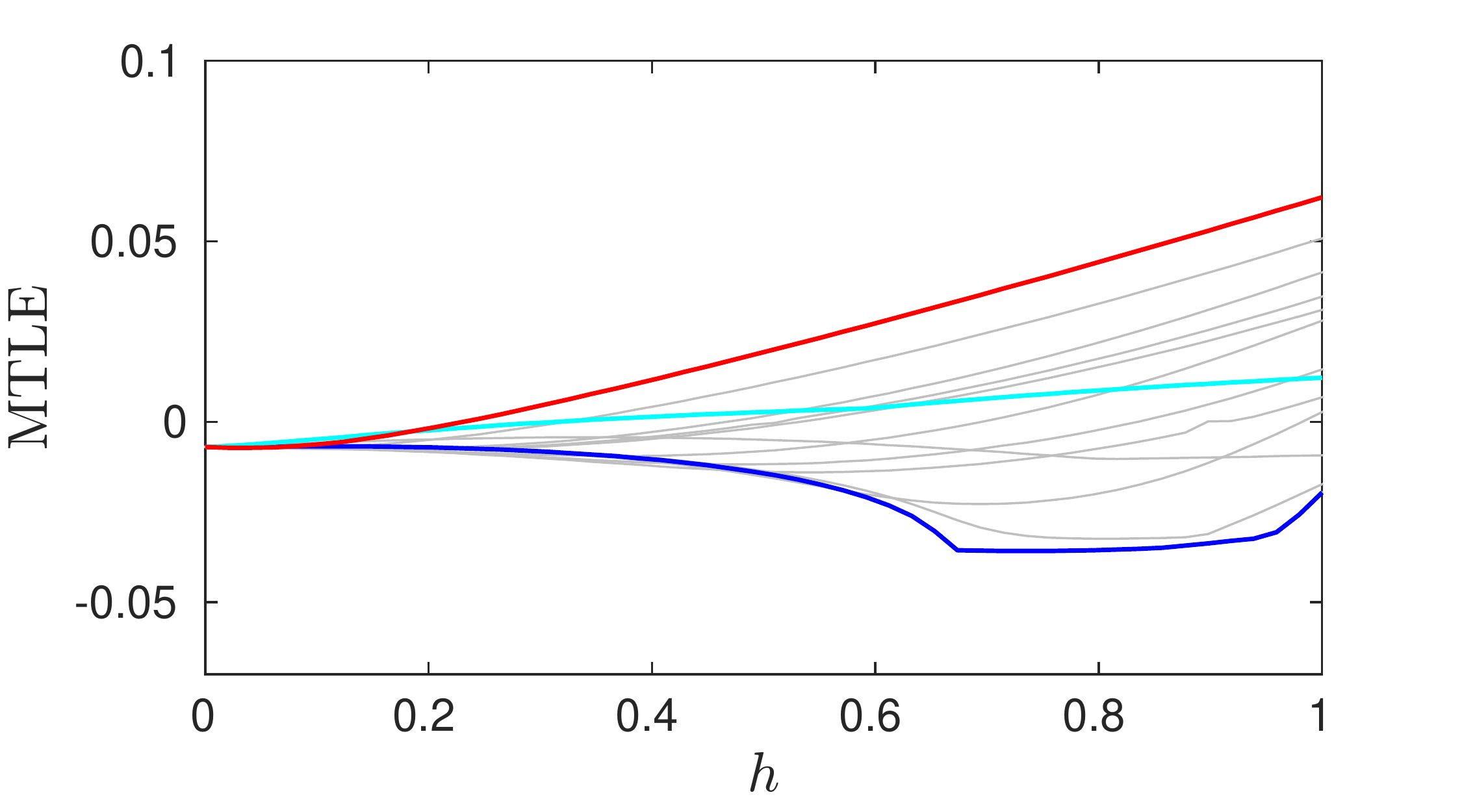}};
    \begin{scope}[x={(image.south east)},y={(image.north west)}]
        \draw[ultra thick,black,dashed] (.15,.48) -- (.4,.48);
        \draw[ultra thick,ForestGreen,dashed] (.4,.48) -- (.9,.48);
        \draw[draw,|-|,ultra thick,ForestGreen] (.42,.95) -- (.905,.95);
        \node at (.66,.99) {\large \textit{AISync}};
        \draw[thick,red] (.6,.85) -- (.66,.85);
        \node at (.69,.85) {$-h$};
        \draw[thick,cyan] (.6,.8) -- (.66,.8);
        \node at (.69,.8) {$+h$};
        \draw[thick,blue] (.6,.75) -- (.66,.75);
        \node at (.69,.75) {$\pm h$};
        \node[anchor=south west,inner sep=0] (image) at (0.16,0.57) {\includegraphics[width=0.27\textwidth]{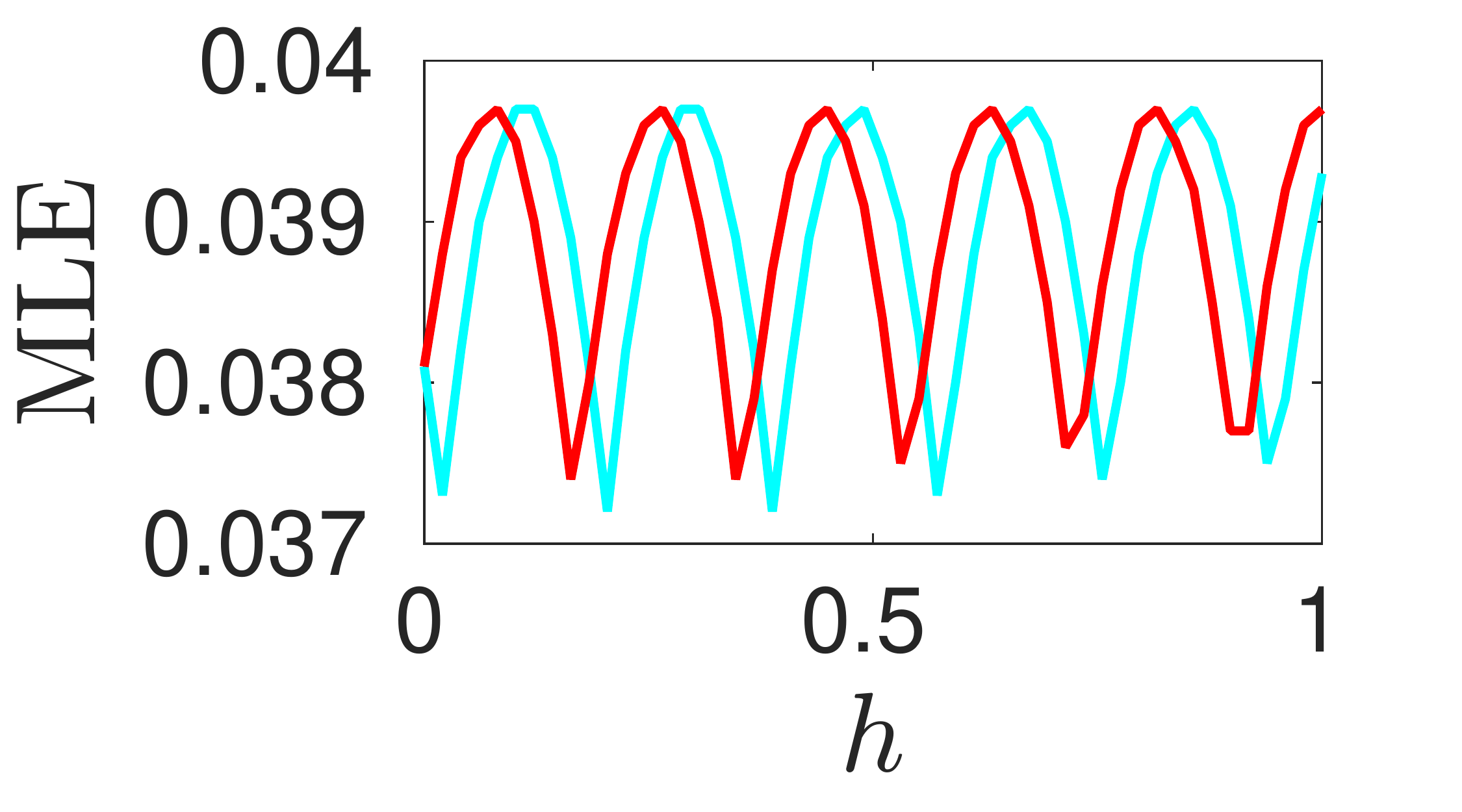}};
        \draw[arc] (.45,.3) to (.55,.35);

        \node[vertex,fill=SkyBlue] (p1) at (.3,.4) {};
        \node[vertex,fill=SkyBlue] (p3) at (.41,.3) {};
        \node[vertex,fill=SkyBlue] (p5) at (.3,.2) {};
        \node[vertex,fill=SkyBlue] (p2) at (.38,.4) {};
        \node[vertex,fill=red] (p4) at (.38,.2) {};
        \node[vertex,fill=red] (p6) at (.27,.3) {};
        \foreach [count=\r] \row in 
        {{0,0,0,0,0,1},
         {1,0,0,0,0,0},
         {0,1,0,0,0,0},
         {0,0,1,0,0,0},
         {0,0,0,1,0,0},
         {0,0,0,0,1,0}}
        {
            \foreach [count=\c] \cell in \row
            {
                \ifnum\cell=1
                    \draw[arc] (p\c) edge (p\r);
                \fi
            }
        }
    \end{scope}
\end{tikzpicture}
}
\caption{Dependence of the MTLE on the heterogeneity $h$ for heterogeneous systems (gray and blue lines) and homogeneous systems ($+h$, cyan line;  $-h$, red line).
A total of $10$ out of $11$ heterogeneous systems exhibit \textit{AISync} for some range of $h\in (0, 1)$.
The most synchronizable heterogeneous system is $\pm h$ (highlighted in blue, and accompanied by its network diagram color-coded by oscillator type). 
The MLE of the amplitude death state is shown as inset for both homogeneous systems. 
The dashed line at zero marks the boundary of linear stability,  and the top green line marks the region $h \in (0.32,1)$ where {\it AISync} occurs. 
}
\label{fig:mtle}
\end{figure}

\begin{figure}[htb]
\subfloat[]{
\begin{tikzpicture}
    \node[anchor=south west,inner sep=0] (image) at (0,0,0) {\includegraphics[width=.95\columnwidth]{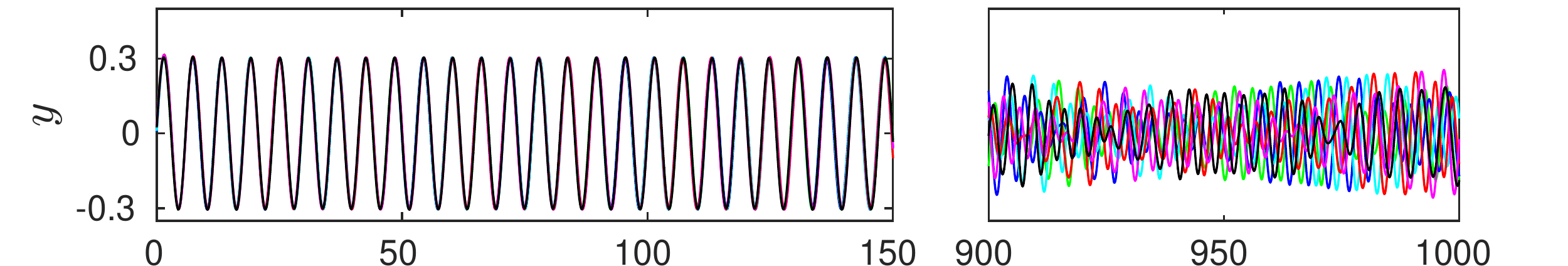}};
    \begin{scope}[x={(image.south east)},y={(image.north west)}]
        \draw[thick,cyan,->] (.1,.82) -- (.92,.82);
        \node[align = left] at (.5,.89) {\footnotesize $+h$};
        \node[] at (.97,.03) {$t$};
    \end{scope}
\end{tikzpicture}
}
\llap{\parbox[b]{5.9in}{(a)\\\rule{0ex}{.8in}}}
\\[-4ex]
\subfloat[]{
\begin{tikzpicture}
    \node[anchor=south west,inner sep=0] (image) at (0,0,0) {\includegraphics[width=.95\columnwidth]{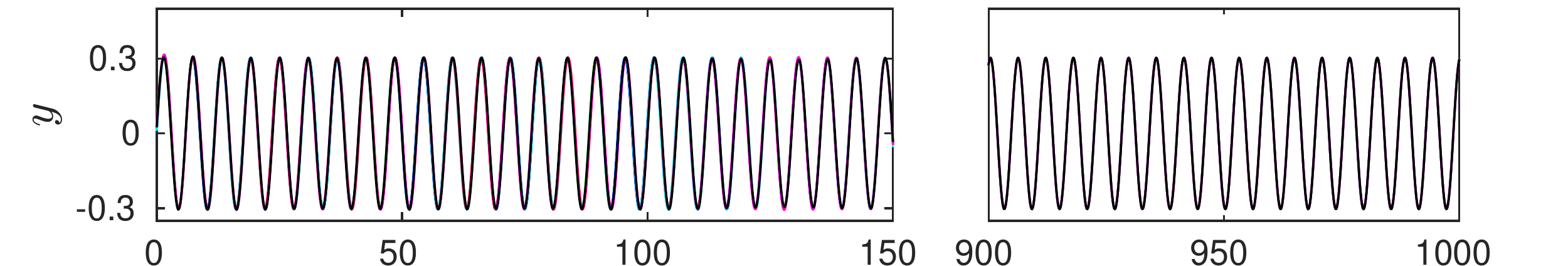}};
    \begin{scope}[x={(image.south east)},y={(image.north west)}]
        \draw[thick,cyan,->] (.1,.82) -- (.41,.82);
        \draw[thick,blue,->] (.42,.82) -- (.92,.82);
        \node[align = left] at (.28,.89) {\footnotesize $+h$};
        \node[align = left] at (.72,.89) {\footnotesize $\pm h$};
        \node[] at (.97,.03) {$t$};
    \end{scope}
\end{tikzpicture}
}
\\[-4ex]
\subfloat[]{
\begin{tikzpicture}
    \node[anchor=south west,inner sep=0] (image) at (0,0,0) {\includegraphics[width=.91\columnwidth]{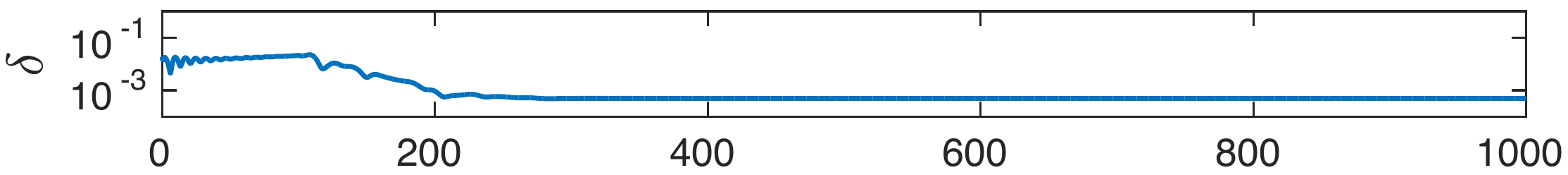}};
    \begin{scope}[x={(image.south east)},y={(image.north west)}]
        \draw[thick,cyan,->] (.105,.73) -- (.19,.73);
        \draw[thick,blue,->] (.20,.73) -- (.96,.73);
        \node[align = left] at (.15,.835) {\footnotesize $+h$};
        \node[align = left] at (.6,.835) {\footnotesize $\pm h$};
        \node[] at (1.01,.1) {$t$};
    \end{scope}
\end{tikzpicture}
}
\\[-2.5ex]
\subfloat[]{
\begin{tikzpicture}
    \node[anchor=south west,inner sep=0] (image) at (0,0,0) {\includegraphics[width=.95\columnwidth]{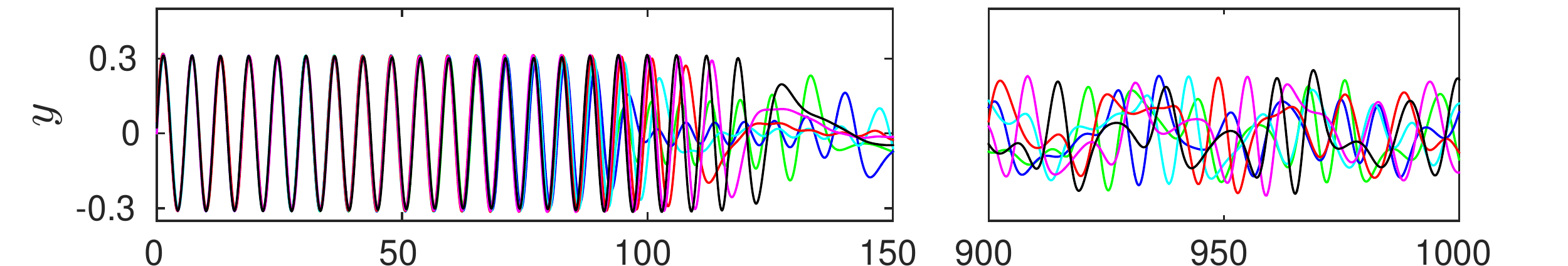}};
    \begin{scope}[x={(image.south east)},y={(image.north west)}]
        \draw[thick,red,->] (.1,.82) -- (.92,.82);
        \node[align = left] at (.5,.89) {\footnotesize $-h$};
        \node[] at (.97,.03) {$t$};
    \end{scope}
\end{tikzpicture}
}
\llap{\parbox[b]{5.9in}{(b)\\\rule{0ex}{.8in}}}
\\[-4ex]
\subfloat[]{
\begin{tikzpicture}
    \node[anchor=south west,inner sep=0] (image) at (0,0,0) {\includegraphics[width=.95\columnwidth]{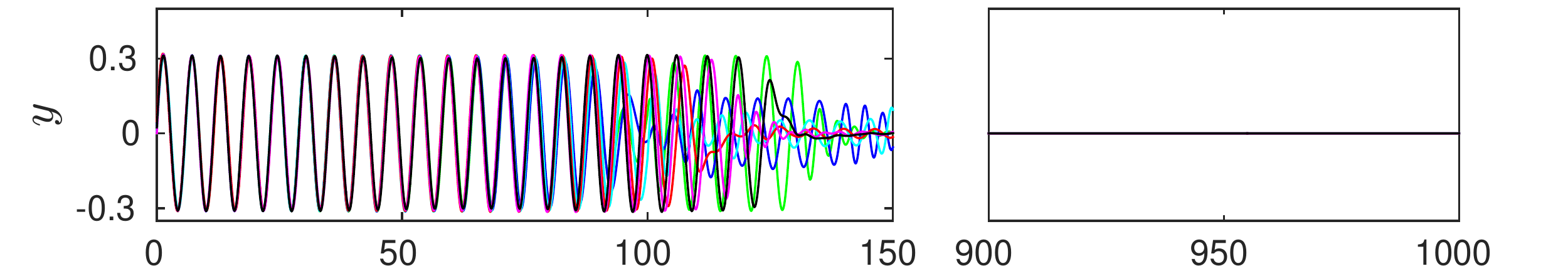}};
    \begin{scope}[x={(image.south east)},y={(image.north west)}]
        \draw[thick,red,->] (.1,.82) -- (.41,.82);
        \draw[thick,blue,->] (.42,.82) -- (.92,.82);
        \node[align = left] at (.28,.89) {\footnotesize $-h$};
        \node[align = left] at (.72,.89) {\footnotesize $\pm h$};
        \node[] at (.97,.03) {$t$};
    \end{scope}
\end{tikzpicture}
}
\vspace{-5mm}
\caption{
(a) Time evolution of typical trajectories of the systems $+h$ and $\pm h$ in figure \ref{fig:mtle} initiated close to the limit cycle synchronous state  $z_j = r_0 e^{i \Omega t}$, where the main panels show the imaginary part of $z_j = x_j + i y_j$, $j = 1,\cdots,6$. 
Top panel: trajectory for the homogeneous system $+h$. 
Middle panel: trajectory starting from the same initial condition when a switch to the heterogeneous system $\pm h$ is performed at $t=100$. 
Bottom panel: time evolution of the synchronization error $\delta$, further 
{\bb demonstrating}
that the trajectory deviates from the limit cycle state for the homogeneous system but converges to it for the heterogeneous one.
(b) Same as in the main panels of (a) but now for the homogeneous system $-h$, where in this case the heterogeneous system $\pm h$ converges to the amplitude death state.
In these examples, the heterogeneity $h$ was chosen to be $0.8$.
{\bb To initiate the system close to the limit cycle synchronous state, we extended $z_j = r_0 e^{i \Omega t}$ backward $\tau$ time units from $t = 0$, and perturbed the components of each oscillator independently at $t = 0$ with a displacement chosen randomly from the interval $(0,0.1)$.}
}
\label{fig:trj}
\end{figure}

Figure~\ref{fig:trj} shows the result of direct simulations of systems initiated close to the limit cycle synchronous state with $h$ set to 0.8. In figure~\ref{fig:trj}(a) we compare the homogeneous system $+h$ with the heterogeneous system $\pm h$. The upper panel trajectory is produced by system $+h$ alone, which loses synchrony over time due to instability. In the lower panel trajectory, $+h$ is switched to the heterogeneous system $\pm h$ at $t = 100$. This stabilizes the synchronous trajectory and no desynchronization is observed for the course of the simulation. As the effect of this switching in the lower panel trajectory is not very visually distinctive, we also plot the synchronization error $\delta$ for 
{\bb $t \in [0,1000]$,}
where $\delta$ is defined as the standard deviation among $z_j$:
\begin{equation}
	\delta = \sqrt{\frac{1}{N}\sum_{j=1}^N \Vert z_j - \bar{z} \Vert^2},\quad \bar{z} = \frac{1}{N}\sum_{j=1}^N z_j.
\end{equation}
The error plot clearly shows that $\delta$ grows for $t<100$ and decreases for $t>100$.
Similarly, in figure~\ref{fig:trj}(b) we compare the homogeneous system $-h$ with the heterogeneous system $\pm h$. The upper panel trajectory (now produced by system $-h$ alone) quickly loses synchrony and evolves into a high-dimensional incoherent state, while the lower panel trajectory converges to the amplitude death state after switching to $\pm h$. Note that this state is different from the limit cycle state in figure~\ref{fig:trj}(a), illustrating that $\pm h$ has two distinct 
{\bb symmetric} states that are stabilized by system 
{\bb asymmetry.}

We further characterize the {\it AISync} property of those systems in terms of the time delay $\tau$. The stability of
{\bb the} homogeneous systems is compared with the heterogeneous system $\pm h$ for a range of $\tau$ and $h$. As shown in figure~\ref{fig:ais}, the parameter space is divided into three regions: a region where the heterogeneous system $\pm h$ and at least one of the homogeneous systems are stable (region I);  a region where the heterogeneous system $\pm h$ is stable but both homogeneous systems are unstable (region II); and a region where all three systems, $\pm h$, $+h$ and $-h$, are unstable (region III). This figure establishes the occurrence of {\it AISync} in the entire region II. We note, moreover, that region II is a conservative estimate of the {\it AISync} region as 
{\bb it}
only 
{\bb considers}
one out of eleven possible heterogeneous systems and  only 
{\bb concerns} the limit cycle synchronous state 
(not accounting for 
{\bb the possibility of a}
stable amplitude death state for $\pm h$; we have verified that this fixed point is unstable for both $+h$ and $-h$ over the entire 
{\bb range of  $h$ and $\tau$}
considered in our simulations). 
%
Another interesting fact to note is that unlike many other delay-coupled systems \cite{cao2013overview},
here larger delay does not always {\bb lead} 
to reduced synchronizability, as both region I and the union of regions I and II expand with increasing time delay $\tau$.
{\bb
This adds to the few existing examples showing time-delay enhanced synchronization \cite{atay2004delays,ernst1995synchronization,dhamala2004enhancement,wang2011synchronous}, which can have implications for phenomena such as the remote synchronization between neurons in distant cortical areas.
}

\begin{figure}[htb]
\centering
\subfloat[]{
\begin{tikzpicture}
    \node[anchor=south west,inner sep=0] (image) at (0,0,0) {\includegraphics[width=.8\columnwidth]{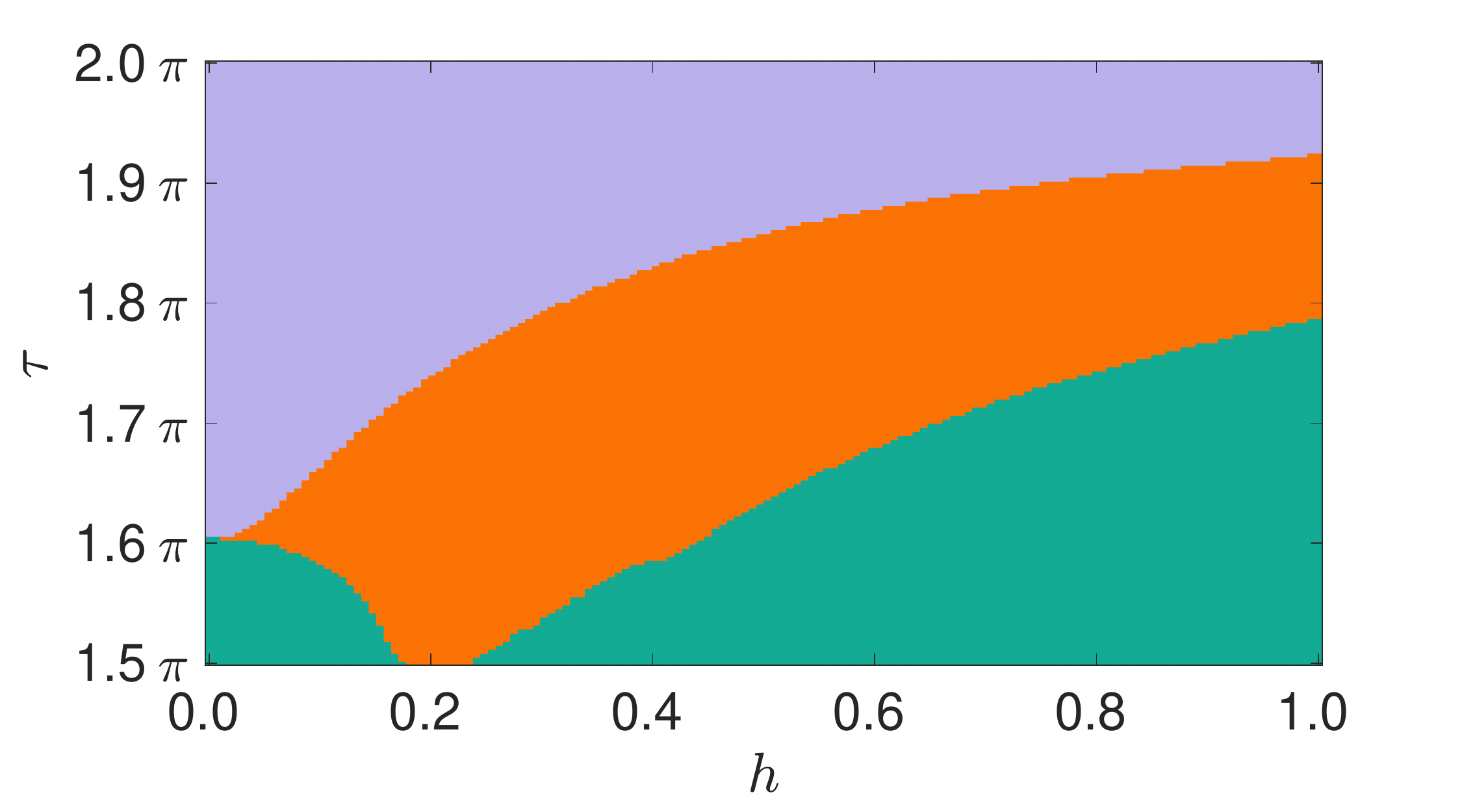}};
    \begin{scope}[x={(image.south east)},y={(image.north west)}]
        \node at (.3,.7) {I};
        \node at (.5,.55) {II};
        \node at (.7,.35) {III};
    \end{scope}
\end{tikzpicture}
}
\vspace{-5mm}
\caption{Regions in the $\tau$ versus $h$ plane categorized by the stability of the limit cycle synchronous state for the $6$-node directed ring network in figure~\ref{fig:mtle}. In region I (purple), the synchronous state is stable for both the heterogeneous system $\pm h$ and at least one homogeneous systems ($- h$ and/or $+ h$). In region II (orange), the synchronous state is stable only for the heterogeneous system. Region II is thus part of the {\it AISync} region. In region III (green), the synchronous state is unstable for the two homogeneous systems as well as for the  heterogeneous one.}
\label{fig:ais}
\end{figure}
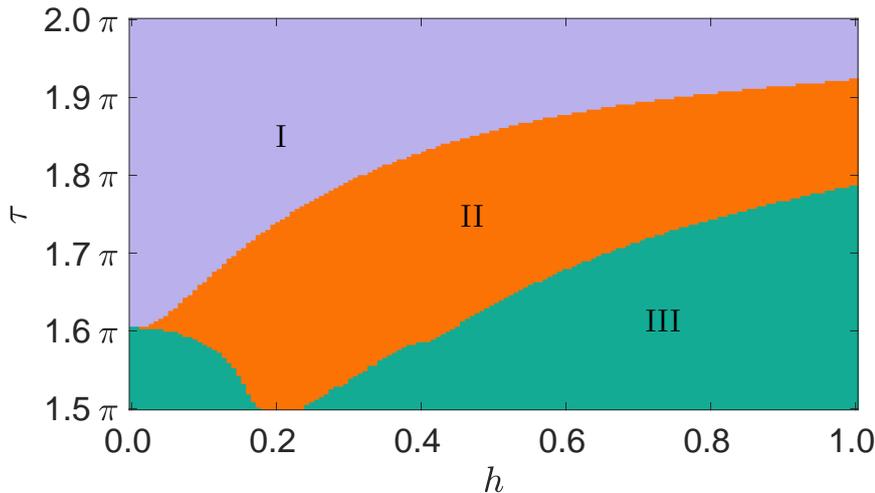

Thus far we have focused on Stuart-Landau oscillators coupled through a directed ring network, which is a directed version of a circulant graph.
Such graphs have the property of admitting a circulant matrix as their adjacency matrix.
Networks of identically coupled oscillators can be much more complex, as most symmetric networks are non-circulant. It is thus natural to ask whether {\it AISync} can be observed for Stuart-Landau oscillators coupled through such {\bb networks.} 

In figure~\ref{fig:ais-1} we show the example of a (weighted) crown network of 8 nodes, which, when weighted, is the  smallest non-circulant vertex-transitive graph \cite{biggs1993algebraic}. In this illustration, the inner edges have weight $a$ and the outer edges have fixed weight $1$. It is intuitively clear that all nodes are identically coupled. 
{\bb This can be verified in  figure~\ref{fig:ais-1}(a)}
by applying 90-degree rotations (which 
{\bb connect nodes of the}
same color) and reflections with respect to the dashed line (which connects nodes of different colors) to show that all nodes belong to the same orbit under automorphisms of the network.

We calculate the stability of the limit cycle synchronous state for both homogeneous systems (all nodes equipped with $\bm{F}^{(1)}$ or all nodes equipped with $\bm{F}^{(2)}$) and a representative heterogeneous system. The heterogeneous system has $\bm{F}^{(1)}$ and $\bm{F}^{(2)}$ arranged alternatingly, as indicated by the colors in figure~\ref{fig:ais-1}(a), which is {\bb a configuration} described by $\bm{D}^{(1)} = \text{diag}\{1,0,1,0,1,0,1,0\}$ and $\bm{D}^{(2)} = \text{diag}\{0,1,0,1,0,1,0,1\}$. 
{\bb These matrices and the adjacency matrix of the
crown network can be simultaneously block diagonalized}
into a block  structure composed of four $2 \times 2$ blocks, which significantly simplifies the calculation of {\bb the} MTLE. The results are shown in figure~\ref{fig:ais-1}(b) for a range of inner edge weight $a$ and oscillator heterogeneity $h$, where the regions I, II, {\bb and} III are defined as before. 
We have also verified {\bb that,} for the entire range of parameters in {\bb figure \ref{fig:ais-1}(b),} the amplitude death state is unstable for both homogeneous systems.
For the same reasons as in figure~\ref{fig:ais}, region II is a conservative estimate of the {\it AISync} region. The {\it AISync} region extends from $a=0$, where the network is an unweighted ring, to  $a=1$, where the network is an unweighted crown.  Incidentally, this example  shows that {\it AISync} can also occur for undirected networks {\bb (previous examples, both in this article and in the literature \cite{PhysRevLett.117.114101,zhang2017asymmetry}, were limited to directed networks).}

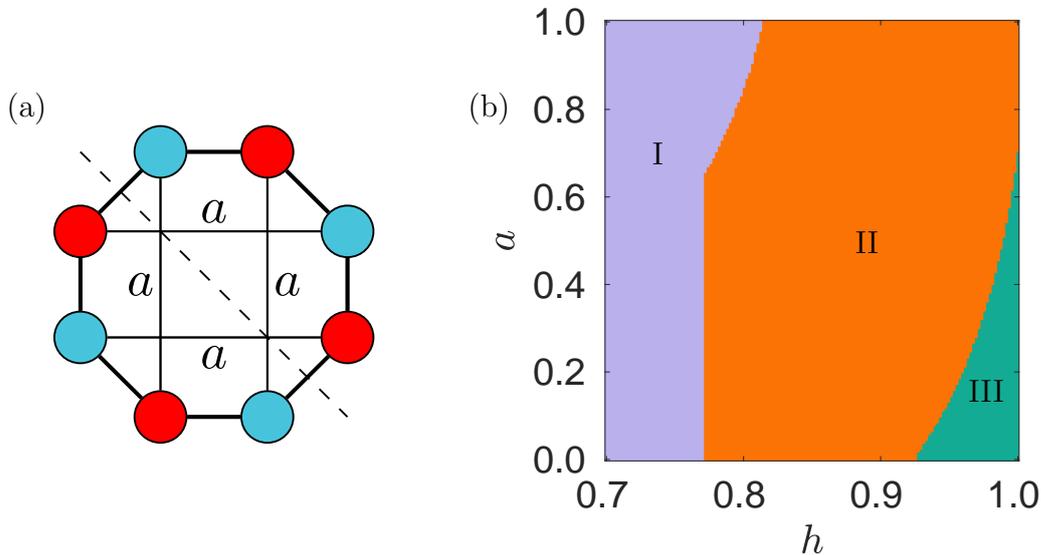
\begin{figure}[htb]
\centering
\subfloat[]{
\resizebox{.3\textwidth}{!}{
\begin{tikzpicture} 
    [vertex/.style={draw,circle,thin}]
    \node[vertex,fill=SkyBlue] (p1) at (2.6,4) {};
    \node[vertex,fill=red] (p2) at (3.4,4) {};
    \node[vertex,fill=SkyBlue] (p3) at (4,3.4) {};
    \node[vertex,fill=red] (p4) at (4,2.6) {};
    \node[vertex,fill=SkyBlue] (p5) at (3.4,2) {};
    \node[vertex,fill=red] (p6) at (2.6,2) {};
    \node[vertex,fill=SkyBlue] (p7) at (2,2.6) {};
    \node[vertex,fill=red] (p8) at (2,3.4) {};
    \node at (3,3.55) {\small $a$};
    \node at (3,2.45) {\small $a$};
    \node at (2.45,3) {\small $a$};
    \node at (3.55,3) {\small $a$};
    \node at (3,1) {};
    \foreach [count=\r] \row in 
    {{0,1,0,0,0,2,0,1},
     {1,0,1,0,2,0,0,0},
     {0,1,0,1,0,0,0,2},
     {0,0,1,0,1,0,2,0},
     {0,2,0,1,0,1,0,0},
     {2,0,0,0,1,0,1,0},
     {0,0,0,2,0,1,0,1},
     {1,0,2,0,0,0,1,0}}
    {
        \foreach [count=\c] \cell in \row
        {
            \ifnum\cell=1
                \draw[thick] (p\c) edge (p\r);
            \fi
            \ifnum\cell=2
                \draw (p\c) edge (p\r);
            \fi
        }
    }
    \draw[dashed] (2,4) edge (4,2);
\end{tikzpicture}
}}
\llap{\parbox[b]{2in}{(a)\\\rule{0ex}{2.3in}}}
\hfil
\subfloat[]{
\begin{tikzpicture}
    \node[anchor=south west,inner sep=0] (image) at (0,0,0) {\includegraphics[width=.5\columnwidth]{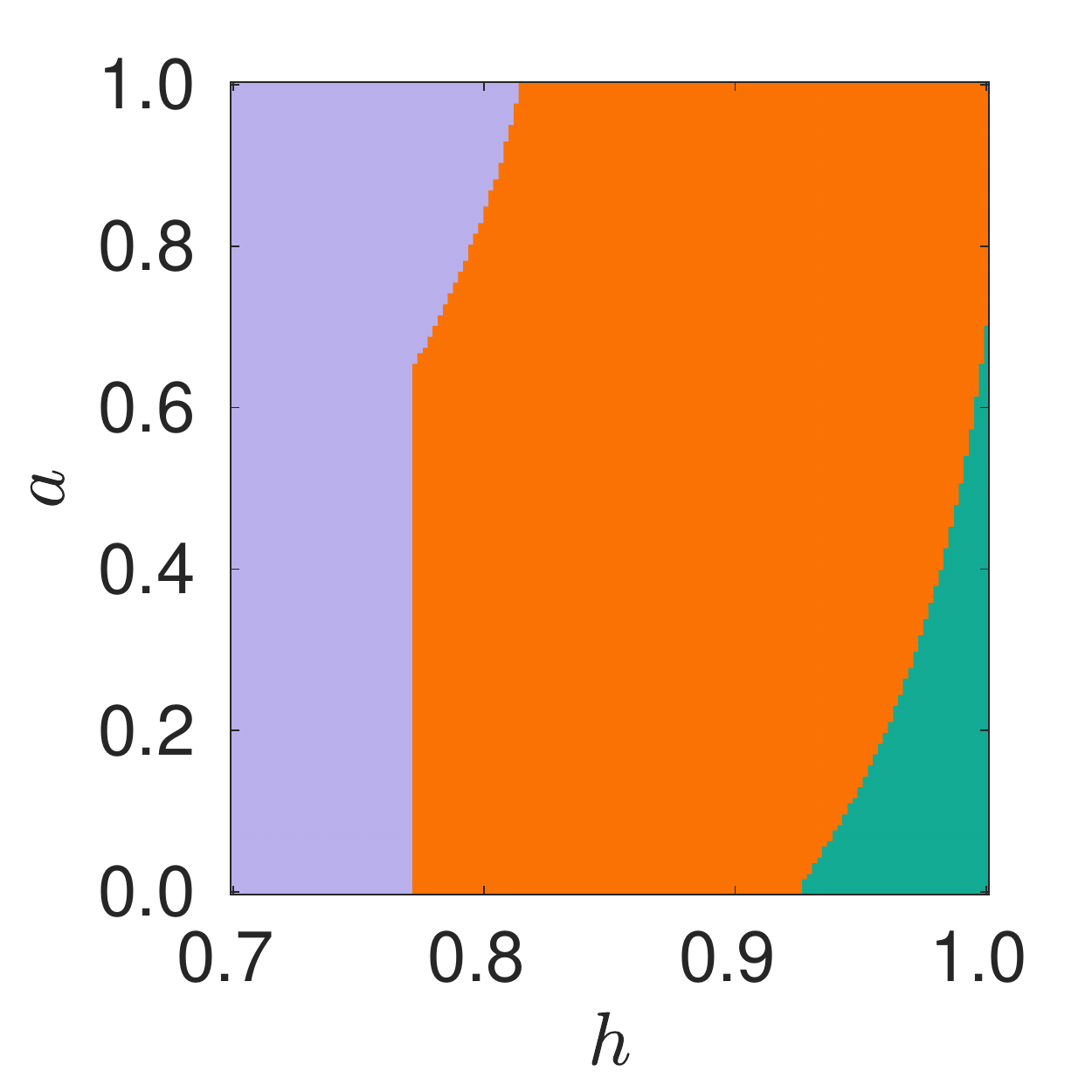}};
    \begin{scope}[x={(image.south east)},y={(image.north west)}]
        \node at (.3,.7) {I};
        \node at (.65,.55) {II};
        \node at (.85,.3) {III};
    \end{scope}
\end{tikzpicture}
}
\llap{\parbox[b]{3.2in}{(b)\\\rule{0ex}{2.3in}}}
\caption{{\it AISync} in a symmetric non-circulant network.
(a) 
{\bb Crown} network of $N=8$ {\bb nodes,} for outer edges of weight 1 and inner edges of weight $a$.
(b) Illustration in the $a$ versus $h$ plane of  the regions categorized by the stability of the limit cycle 
synchronous state, where 
the heterogeneous
system alternates the two types of oscillators as color-coded in (a). As in the previous example, region II is part of the {\it AISync} region,
{\bb since in this region synchronization is unstable for both  homogeneous systems ($+h$ and $-h$) but is stable for the heterogeneous one.}  
}
\label{fig:ais-1}
\end{figure}

\subsection{Synchronization in non-symmetric networks induced by oscillator heterogeneity}
\label{sec:nonsym} 

The scenario in which identically coupled oscillators synchronize stably only when the oscillators themselves are nonidentical is, arguably, {\it AISync} in its 
{\bb most compelling}
form, as all the heterogeneity can then be attributed to the oscillators and there is no potential for compensatory 
heterogeneity {\bb to result}
from the network structure.
But
{\bb the possibility of synchronization induced by}
oscillator heterogeneity 
is not restricted to 
{\bb such symmetric networks,}
{\bb and we hypothesize that}
this effect can be prevalent {\bb also} in networks that do not have symmetric structure. 
{\bb In examining this hypothesis, it is useful to note that our}
analysis extends naturally to networks that have the same (weighted) indegree for all nodes {\bb (as defined in equation \eqref{balance})} 
but are otherwise arbitrary. 

Figure~\ref{fig:ais-2} shows an example for a $6$-node non-symmetric network of Stuart-Landau oscillators. The network, which is directed and 
{\bb has}
both positive and negative edge weights, 
{\bb is composed of}
three symmetry 
{\bb clusters,} as defined by the orbits of its automorphism group. Each symmetry cluster consists of two nodes that are diagonally opposite and  can be mapped to each other by rotations of 
{\bb 180 degrees}
in the representation of figure~\ref{fig:ais-2}(a). Thus, the oscillators are indeed not identically coupled. As a representative heterogeneous system to be compared with the homogeneous systems {\bb ($+h$ and $-h$),} 
we consider four nodes equipped with the dynamics $\bm{F}^{(1)}$ and the other two nodes with $\bm{F}^{(2)}$ (as indicated by the colors in figure~\ref{fig:ais-2}(a), {\bb where $\bm{F}^{(1)}$ and $\bm{F}^{(2)}$ are defined as in  Sec.~\ref{sec:nind}).} In figure~\ref{fig:ais-2}(b) we show the results of the stability analysis for a range of {\bb the} effective coupling strength $\sigma\mu$ and oscillator heterogeneity $h$, with the same definition of regions I, II and III as 
above. 
{\bb Again,} we verified that, for the 
range of parameters in figure \ref{fig:ais-2}(b), 
the amplitude death state is unstable for both homogeneous systems. Region II occupies a sizable portion of the diagram, showing that the scenario in which the oscillators are required to be 
{\bb nonidentical} 
for the network to synchronize stably can also be common for non-symmetric networks.

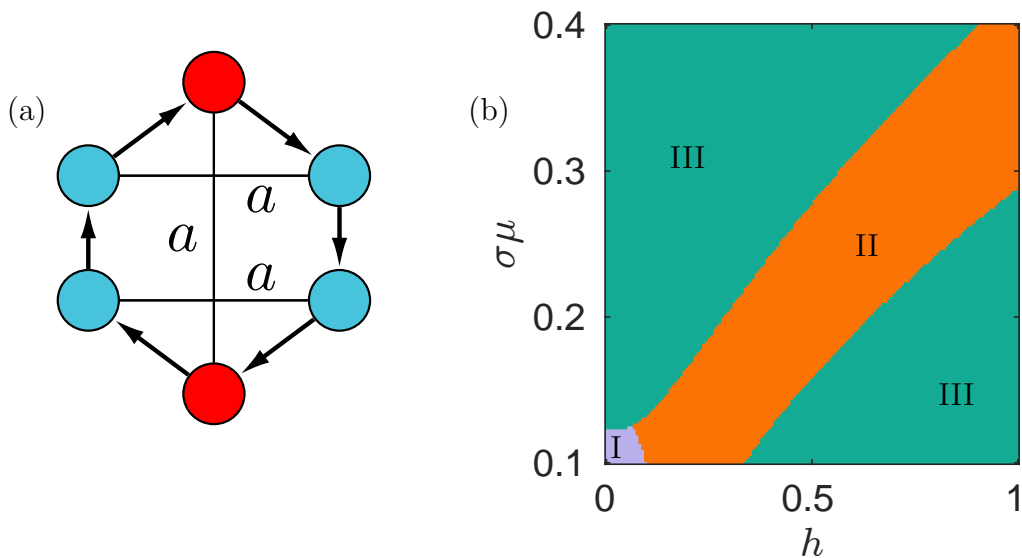
\begin{figure}[htb]
\centering
\subfloat[]{
\resizebox{.3\textwidth}{!}{
\begin{tikzpicture} 
	[vertex/.style={draw,circle,thin},
    arc/.style={draw,-{Latex[length=2mm, width=1mm]}}]
    \node[vertex,fill=red] (p1) at (3,4) {};
    \node[vertex,fill=SkyBlue] (p2) at (3.8,3.4) {};
    \node[vertex,fill=SkyBlue] (p3) at (3.8,2.6) {};
    \node[vertex,fill=red] (p4) at (3,2) {};
    \node[vertex,fill=SkyBlue] (p5) at (2.2,2.6) {};
    \node[vertex,fill=SkyBlue] (p6) at (2.2,3.4) {};
    \node at (3.3,3.25) {\small $a$};
    \node at (3.3,2.75) {\small $a$};
    \node at (2.8,3) {\small $a$};
    \node at (3,1) {};
    \foreach [count=\r] \row in 
    {{0,0,0,2,0,1},
     {1,0,0,0,0,2},
     {0,1,0,0,2,0},
     {2,0,1,0,0,0},
     {0,0,2,1,0,0},
     {0,2,0,0,1,0}}
    {
        \foreach [count=\c] \cell in \row
        {
            \ifnum\cell=1
                \draw[arc,thick] (p\c) edge (p\r);
            \fi
            \ifnum\cell=2
                \draw (p\c) edge (p\r);
            \fi
        }
    }
\end{tikzpicture}
}}
\llap{\parbox[b]{2in}{(a)\\\rule{0ex}{2.3in}}}
\hfil
\subfloat[]{
\begin{tikzpicture}
    \node[anchor=south west,inner sep=0] (image) at (0,0,0) {\includegraphics[width=.5\columnwidth]{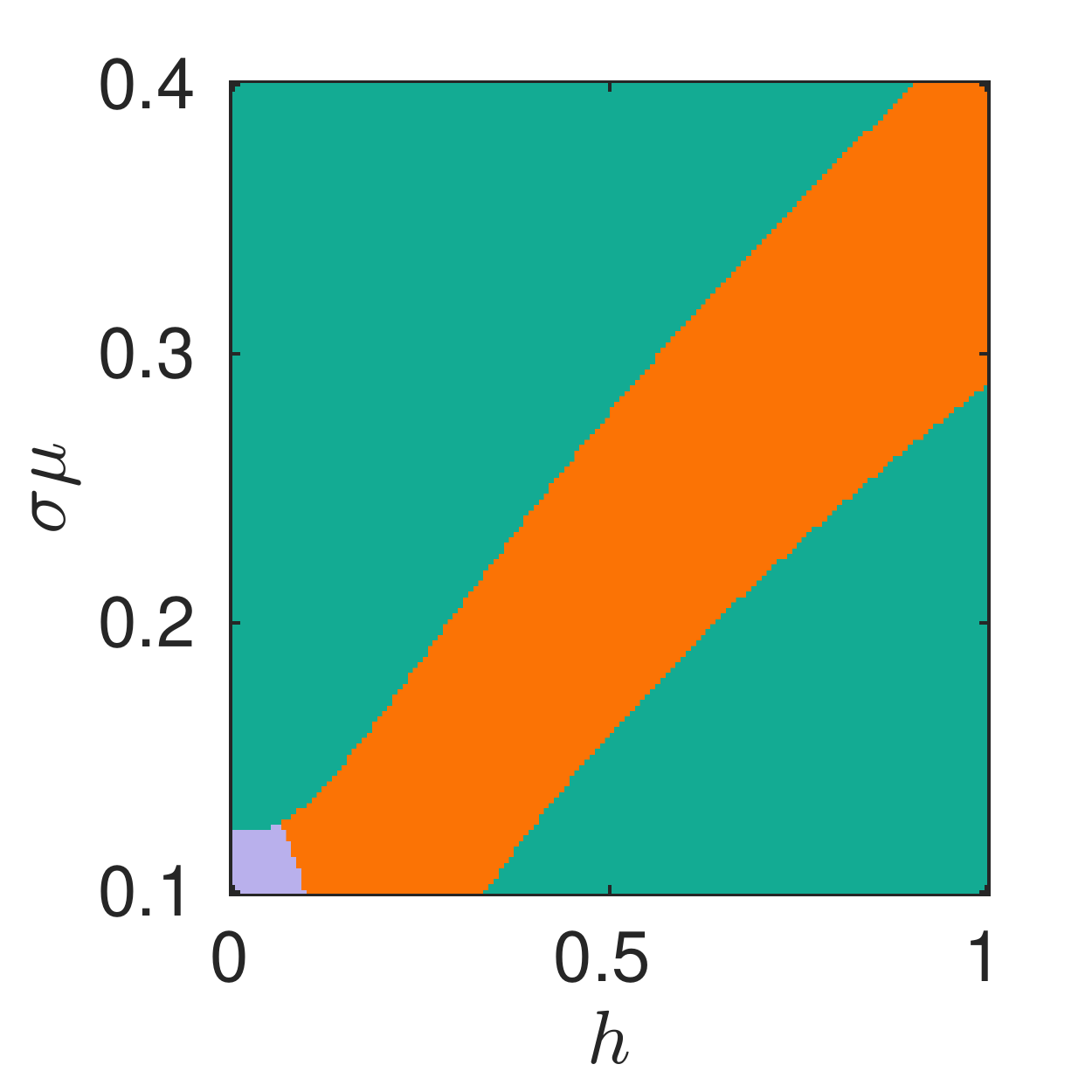}};
    \begin{scope}[x={(image.south east)},y={(image.north west)}]
        \node at (.35,.7) {III};
        \node at (.65,.55) {II};
        \node at (.8,.3) {III};
        \node at (.23,.21) {I};
    \end{scope}
\end{tikzpicture}
}
\llap{\parbox[b]{3.2in}{(b)\\\rule{0ex}{2.3in}}}
\caption{Synchronization induced by oscillator heterogeneity in a non-symmetric network.
(a) Network of $N=6$ nodes with outer (directed) edges of weight $+1$ and inner (undirected) edges of weight $a=-0.1$, which has $3$ symmetry clusters.
(b) Illustration in the $\sigma \mu$ versus $h$ plane of the regions categorized by the stability of the limit cycle synchronous state, where the heterogeneous
system has two oscillators of one type and four oscillators of the other type, as indicated by the colors in (a).  Once again, region II corresponds to the scenario
in which the homogeneous systems $-h$ and $+h$ do not synchronize stably whereas the heterogeneous system does.}
\label{fig:ais-2}
\end{figure}

\subsection{\bb Generalization to unrestricted parameters}
\label{sec:genr} 

{\bb
We 
note that our Stuart-Landau system exhibits even
stronger notions of {\it AISync} than the one  considered thus far.  
For example, if a mixture of heterogeneous oscillators with parameters $+h$ and $-h$ can synchronize stably, 
then 
it may be natural to ask whether the homogeneous systems are all unstable even when the oscillator parameter is chosen
from the entire interval $[-h,+h]$, rather than just from the pair $\{-h, +h\}$. 
The system in equation~\eqref{eq:1} exhibits {\it AISync} 
also in this stronger sense.
The region between $\tau = 1.5\pi$ and $\tau = 1.6\pi$ in figure~\ref{fig:ais} provides one such example, 
where both homogeneous systems $+h$ and $-h$ are unstable for any $h \in [0,1]$, while the 
heterogeneous system $\pm h$ is stable for a range of $h$ in this interval. 
A similar result holds for heterogeneity-induced synchronization in non-symmetric networks, as illustrated by 
the example in figure~\ref{fig:ais-2} for $\sigma\mu$ larger than~$0.13$.

One can further ask whether it is possible for a heterogeneous system to be stable for some $h$ 
while the homogeneous systems are unstable for any $h\in (-\infty,+\infty)$.
We have extended the range of $h$ in figure~\ref{fig:ais} to verify numerically that no homogeneous system can synchronize for any $h$ in the region $1.5\pi < \tau < 1.6\pi$
(and, similarly, in figure~\ref{fig:ais-2} for any $h$ in the region $\sigma\mu > 0.13$).
Thus, these are scenarios in which the oscillators need to be nonidentical  for the system to synchronize
even if the homogeneous system has unrestricted access to the values of the parameter $h$.

Naturally, we can also imagine heterogeneous systems in which each oscillator is allowed to take an independent 
value of the parameter  $h$  in the full 
interval~$(-\infty,+\infty)$. Computational challenges aside, this would only show that the phenomenon of {\it AISync} is even more common, since there would 
then be a larger set of heterogeneous {\bb systems to choose from.}
}

\section{Final remarks}
\label{sec:concl}

Motivated by the recent discovery of {\it asymmetry-induced synchronization}, here we 
established a general stability analysis method for 
{\bb demonstrating and examining identical synchronization among}  
nonidentical oscillators.\   
This can be seen as a generalization of the 
{\bb standard}
master stability analysis 
{\bb to non-perturbative regimes of parameter mismatches, and is illustrated for {\bb systems with}
Laplacian- and adjacency-matrix coupling 
as well as time-delay coupling.}
In establishing our formalism, 
{\bb we first characterized the most general conditions under which nonidentical oscillators can synchronize completely for the various coupling schemes. 
When the coupling is non-diffusive, the balanced input conditions required for the synchronization of identical oscillators \cite{golubitsky2016rigid,aguiar2011dynamics} is replaced by 
conditions that involve both the node dynamics and the coupling term. We then established our approach to simultaneously block diagonalize the matrices in the variational equation, which reduces dimension and facilitates stability analysis.}

This new framework was 
applied to 
{\bb networks of heterogeneous}
delay-coupled Stuart-Landau oscillators and reveals {\it AISync} as a 
{\bb robust}
behavior.  We identify coupling delay as a new key ingredient leading to {\it AISync} in this class of systems, which suggests that {\it AISync} may be more common than previously anticipated 
{\bb in real systems,}
where delay is ubiquitous {\bb \cite{kolmanovskii2013introduction,niculescu2001delay}.}
{\bb
The possibility of {\it AISync}, along with the conditions we establish for complete synchronization, has the potential to lead 
to new optimization and control approaches 
focused on creating or enhancing synchronization stability 
in networks that may not synchronize spontaneously when the dynamical units are identical. 
By tuning the oscillators to suitable nonidentical parameters, such approaches promise to 
be useful specially when the network structure is fixed, since they could rely solely on nonstructural 
degrees of freedom associated with the local (node) dynamics.
}


Our analysis of the heterogeneous Stuart-Landau system 
is far from exhaustive, as we  explored only slices of 
{\bb its}
vast parameter space and our parameter choices were not 
{\bb fine-tuned.}
In particular, we  only considered real coupling strength $\sigma$, while complex coupling strength (known as conjugate coupling) has been shown to have significant impact on stability \cite{choe2010controlling}; the effect of the oscillation amplitude parameter $\lambda$, not varied here, 
{\bb also warrants}
further investigation. Moreover, we focused mainly on heterogeneous systems of only two kinds of oscillators,  
while more diverse populations of oscillators are yet to be explored in detail. 
Another promising future direction 
{\bb concerns}
the underlying network structure, as it is still an open question whether there exist ways to characterize a system's potential to exhibit {\it AISync} 
on the basis of its network structure. In particular, it would be desirable to identify algebraic indexes determined by the coupling matrix that could 
provide such {\bb a} 
characterization (similar to the way matrix eigenvalues determine synchronizability of certain coupled-oscillator systems).

{\bb Finally, we emphasize that while we have explicitly developed our framework for three widely adopted forms of coupling, complete synchronization can be investigated through 
a similar formulation whenever necessary and sufficient conditions analogous to those in  Table~\ref{tbl:tbl1} can be derived. 
In particular, this may include systems with different types of coupling matrices, other forms of time delay, heterogeneous interaction functions, and explicit periodic time dependence in the coupling terms.  
Another area for future research concerns extending this work to systems in which the  synchronization applies to some (but not all) dynamical variables or to functions of the variables (but not the variables themselves), as  found in various applications.  
Such systems are candidates to exhibit yet new synchronization phenomena,  which we hope will be revealed by future research.
}

\ack{This work was supported by ARO Grant No. W911NF-15-1-0272.}

\section*{References}
\bibliographystyle{nar}
\bibliography{net_dyn}

\end{document}